\documentclass[12pt]{iopart}
\usepackage{comment} 
\usepackage{amssymb} 
\usepackage{graphicx} 
\usepackage[latin1]{inputenc} 
\usepackage{subfigure} 
\usepackage{rotating} 
\usepackage{multirow} 

\newcommand{\eqref}[1]{(~\ref{#1})} 
\newcommand{\sign}{\mathrm{sign}\,} 
 
\usepackage{url,hyperref} 
\usepackage{dsfont} 
 
\begin{document} 
 
\title{Chiral Majorana edge states in HgTe quantum wells} 
 
\author{L. Weithofer$^1$, P. Recher$^1$}
\address{$^1$ Institute for Mathematical Physics, TU Braunschweig,\\
38106 Braunschweig, Germany} 
\ead{p.recher@tu-bs.de}
\begin{abstract} 
HgTe-based quantum wells (QWs) recently attracted a lot of attention for the realization of a two-dimensional topological insulator with protected helical edge states. 
Another class of topological systems are topological superconductors (TSCs) with Majorana edge states. In this paper, we show how proximity induced s-wave superconductivity in the bulk of HgTe-QWs and in the presence of a Zeeman field can exhibit a TSC with chiral Majorana edge states. We calculate the topological invariants and the corresponding Majorana edge states explicitly within a four-band model accounting for inversion symmetry breaking terms due to Rashba spin-orbit coupling and bulk inversion asymmetry present in these QWs.
\end{abstract} 
\submitto{\NJP}
\maketitle
\section{Introduction}
The search for Majorana quasiparticles in condensed matter systems is a subject of great fundamental and practical interest. Majorana quasiparticles have been recently proposed to exist in various systems which typically combine strong spin-orbit
coupling (SOC), the proximity to an s-wave superconductor and magnetic
elements~\cite{Hasan2010, Qi2011, Alicea2012, Beenakker2013}. 

In particular, Majorana bound states (MBS) have been proposed to emerge in vortex cores on the surface of 
strong three-dimensional (3D) topological insulators (TIs)~{\cite{Fu2008}} or at
the edges of two-dimensional TIs when brought to the
proximity of an s-wave superconductor and a
magnet~{\cite{Fu2009, Nilsson2008}}. Characteristic features of Majorana fermions have also
been predicted in spin-orbit coupled two-dimensional semiconductors~\cite{Sau2010, Alicea2010a} and one-dimensional spin-orbit coupled semiconductor nanowires~\cite{Oreg2010}, and very recent experiments show
signatures of MBS which are currently being scrutinized~\cite{Mourik2012,
  Deng2012, Das2012}. In these semiconductor systems, time-reversal
  symmetry is broken by a Zeeman splitting which, in addition to the proximity
  of an s-wave superconductor, is necessary to induce a TSC. In such systems, a large SOC is desireable as the vulnerability of the energy gap to disorder typically
depends on the size of SOC~\cite{Potter2011, Sau2012}. If the
Zeeman splitting is induced by external magnetic fields, low field strengths and therefore small orbital effects are beneficial for the superconducting correlations, which favors semiconducting materials with large effective g-factors.\par
In this work we show that a doped HgTe-QW can implement a TSC, when s-wave
superconductivity and a Zeeman-field is induced in the QW. In order to
  describe the HgTe-QW, we use the four-band BHZ model~\cite{Bernevig2006,
    Rothe2010, Koenig2008} valid near the transition between the inverted and
  non-inverted regime. In the inverted regime and in the absence of superconductivity and a Zeeman gap, the well-known helical edge states appear in the spectral bulk gap \cite{Bernevig2006, Koenig2007, Roth2008}. Importantly, HgTe-QWs are intrinsically subjected to inversion symmetry breaking SOC due to bulk inversion asymmetry (BIA)~\cite{Koenig2008, Winkler2012} which breaks the Kramer's spin-degeneracy. We show that BIA-SOC alone is sufficient to induce a TSC when superconductivity and a Zeeman gap are added. However, tunable (extrinsic) inversion symmetry breaking terms provided by Rashba-SOC can be added as well.

It is known that HgTe-QWs are characterized by a very large Rashba-SOC~\cite{Gui2004}, which can be one order of magnitude larger compared to other III-V compound semiconductors~\cite{Gui2004, Heida1998}. In~\cite{Gui2004}, Rashba-splittings in inverted HgTe-QWs as large as \mbox{30\,meV} were reported. In addition, the effective Zeeman gap in a perpendicular magnetic field was recently calculated and measured in zero gap samples to be consistent with an effective g-factor $g^{*}\sim 55.5$~\cite{Buettner2010}. This shows that HgTe-based QWs could be a feasible candidate for a proximity induced TSC. 

In addition to a potentially feasible implementation of already existing ideas, we show that the combination of BIA and Rashba SOC provides an interesting handle on the topological properties of the TSC. 
In particular, we show that (consistent with topological invariants) the
chirality of Majorana edge states can be reversed by changing the relative
strength of the two SOC terms. \par
In contrast to the proposal of using
1D helical edge states as a platform for MBS~\cite{Fu2008}, we would
like to point out in this work the possibility of switching between chiral (Majorana) and
the well-known helical edge states on the same one-dimensional boundary of a
HgTe-QW. \par
The paper is organized as follows. In~\sref{sec:Model} we introduce the setup
and the relevant model Hamiltonian. Subsequently, we describe the bulk
properties of the system, emphasizing the topologically non-trivial and
trivial phases of the system in~\sref{sec:bulk}. In~\sref{sec:Edge}, we solve
the chiral Majorana edge states at a
domain wall between topologically non-trivial and trivial phases and calculate
the group velocity along the edge. Finally, we compare the chiral Majorana edge states
that emerge when the system is in a TSC phase to the helical edge states that
emerge in a TI phase in~\sref{sec:Comparison}. 
 
\section{The Model} \label{sec:Model}
\begin{figure*}[hbtp]
\begin{center}
\includegraphics[width=0.3\textwidth]{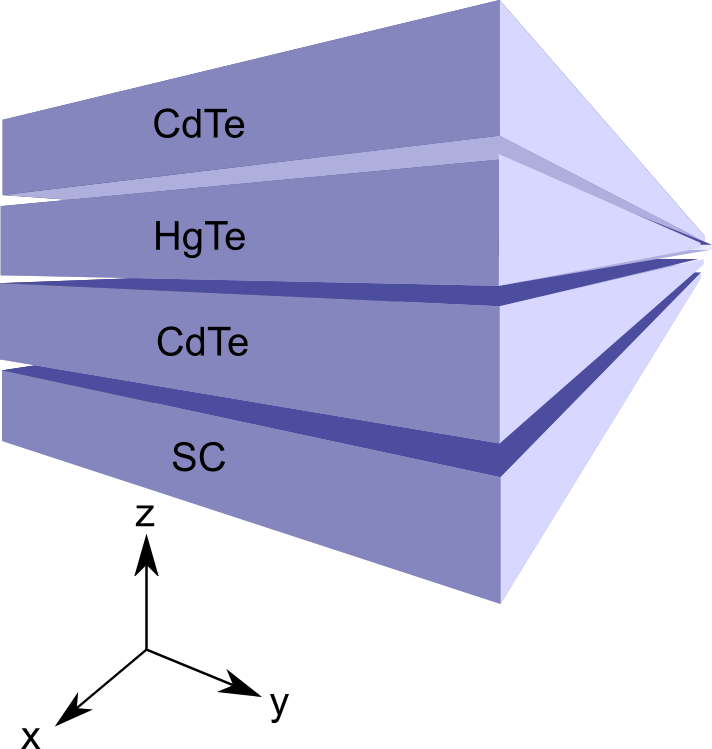}
\caption{HgTe-QW in proximity to an s-wave bulk superconductor (SC).}
\label{fig:TI_majorana_setup}
\end{center}
\end{figure*}

We consider a doped HgTe-QW in the $xy$-plane in contact to an s-wave
superconductor, as sketched
in~\fref{fig:TI_majorana_setup}. There are two inversion\footnote{i.\,e.
$(x,y,z)\rightarrow-(x,y,z)$}-asymmetry induced SOC
effects: Firstly, BIA-induced SOC is intrinsically present as a consequence of the zinc
blende structure of the underlying crystal. Secondly, so-called Rashba-SOC arises when an external electric field is
applied in z-direction. In addition, a Zeeman field is present in $z$-direction which can be induced by the proximity to a ferromagnetic insulator (via spin-exchange interaction), or intrinsically by using Mn-doped (Hg$_{1-y}$Mn$_y$Te) QWs and a small external magnetic field polarising the Mn-spins~\cite{Liu2008b}. Although a magnetic field induces also an orbital effect not taken into account here, we note that the vector potential of a perpendicular magnetic field induces an additional Zeeman splitting between the Kramer's spins with an effective giant g-factor of $g^{*}\sim55.5$ \cite{Buettner2010} so that only small external magnetic fields below 1T would be needed.\par  
The resulting QW system is described in terms of the BHZ model~\cite{Bernevig2006}, complemented with extra terms accounting for
BIA~\cite{Koenig2008}, Rashba~\cite{Rothe2010}, Zeeman
field~\cite{Koenig2008, Liu2008b}, and the proximity to an s-wave superconductor~\cite{Guigou2010}. 
The Hamiltonian can be written in the basis order
$\psi_{E1+},\psi_{H1+},\psi_{E1-},\psi_{H1-},
\psi_{E1-}^\dagger,\psi_{H1-}^\dagger,-\psi_{E1+}^\dagger,-\psi_{H1+}^\dagger$ in
terms of products of the Pauli matrices $\underline{s}, \underline{\sigma},
\underline{\tau}$ and the identity matrix, where $\underline{s}$ acts in
pseudospin ($E1,H1$)-space, $\underline{\sigma}$ acts in Kramer's spin space\footnote{The quantum number that transforms under time
  reversal. The electron spin is not a good quantum number in the presence of
  spin-orbit interactions.} and
$\underline{\tau}$ acts in electron-hole space. The extended BHZ-model
Hamiltonian reads
\begin{equation}
\eqalign{\fl
H=\left(\epsilon(\hat{k}^2)+M(\hat{k}^2)s_z+A\hat{k}_xs_x\sigma_z-A\hat{k}_ys_y\right)\tau_z\cr
+h\left(\cos 2\theta \sigma_y+\sin 2\theta \sigma_x\right)s_y\tau_z\cr
+R_0\frac{\left(s_z+1\right)}{2}\left(\hat{k}_x\sigma_y-\hat{k}_y\sigma_x\right)\tau_z\cr
+\left(B_++B_-s_z\right)\sigma_z+\left(\Delta_++\Delta_-s_z\right)\tau_x.}
\label{eq:fullHamiltonian}
\end{equation} 
with 
\begin{eqnarray}
\epsilon(\hat{k}^2)=C-D(\hat{k}_x^2+\hat{k}_y^2),  \qquad &
M(\hat{k}^2)=M-B(\hat{k}_x^2+\hat{k}_y^2)\\
B_\pm=\left(B_E\pm B_H\right)/2,  \qquad & \Delta_\pm=\left(\Delta_E\pm\Delta_H\right)/2 .
\end{eqnarray}
Here, $\hat{k}_x$ and $\hat{k}_y$ denote the differential operators
$-i\partial_x$ and $-i\partial_y$. \par
The material parameters $A, B, D,
M$ depend on the geometry of the QW. In particular, the mass
parameter $M$ is tunable by the QW thickness. The BHZ-model was successfully used to explain experimental transport data for mass-parameters in the range $-24$ meV $\leq M\leq$ 10 meV~\cite{Tkachov2011}.  The parameter $C$ defines the doping of the QW (tunable by top and/or bottom gates). \par 
The two additional SOC terms have coefficients denoted by $h$
for the BIA term and $R_0$ for the linear Rashba term\footnote{Rashba terms
of higher order in $k$ are neglected.}. As the BIA term is not invariant under rotation in the $xy$
plane, the angle $\theta$ is introduced to describe the angle between the $x$-axis and the $[100]$ crystal direction. \par
The pairing terms $\Delta_E$ and $\Delta_H$ describe the effect of the s-wave
superconductor on the $E1$ and $H1$ bands. We neglect a possible off-diagonal
proximity coupling between $E1$ and $H1$ bands and without loss of generality
assume that the pairing terms $\Delta_E$ and $\Delta_H$ in~\eref{eq:fullHamiltonian} are real \cite{Guigou2010}. \par
Typical values of the above mentioned parameters are presented in~\tref{tab:HgTeValues}. As Zeeman energy of the $E1$ and $H1$ bands, we use $B_E=1.3$\,meV and
$B_H=-0.07$\,meV, respectively. In addition, ${\cal E}=3$\,mV/nm is assumed as
an effective electric field invoking the Rashba-SOC. Unless stated
otherwise, we also use a doping parameter of $C=7.8$\,meV and $\theta=0$, as
well as $\Delta_E=0.5$\,meV and $\Delta_H=0.5$\,meV for the superconducting proximity terms\footnote{In reality, $\Delta_E$ and $\Delta_H$ will not be exactly equal, however, in our parameter regime $C\approx -M$ (see below), the dominant contribution will come from $\Delta_E$.}.

\begin{table*}[htbp]
\begin{center}
\begin{tabular}{c|c|c|c|c|c}
$A$              & $B$              &   $D$                & $M$       & $h_{\mathrm{BIA}}$      & $R_0/(e{\cal E})$\\
\hline
\rule{0pt}{3ex}
3.645\,eV\,\r{A} & $-68.6\,\mathrm{eV\r{A}}^2$ & $-51.2\,\mathrm{eV\r{A}}^2$  &$-0.008$\,eV      & $0.0016$\,eV    &$-15.6$\,nm$^2$    
\end{tabular}
\end{center}
\caption{Typical values for the parameters of the inverted HgTe-QW as taken
  from~\cite{Koenig2008, Rothe2010}.}
\label{tab:HgTeValues}
\end{table*}

\section{Bulk Physics}\label{sec:bulk}
\subsection*{BIA-SOC is sufficient}
\begin{figure}[hbtp]
\begin{center}
\subfigure[]{\includegraphics[width=0.3\textwidth]{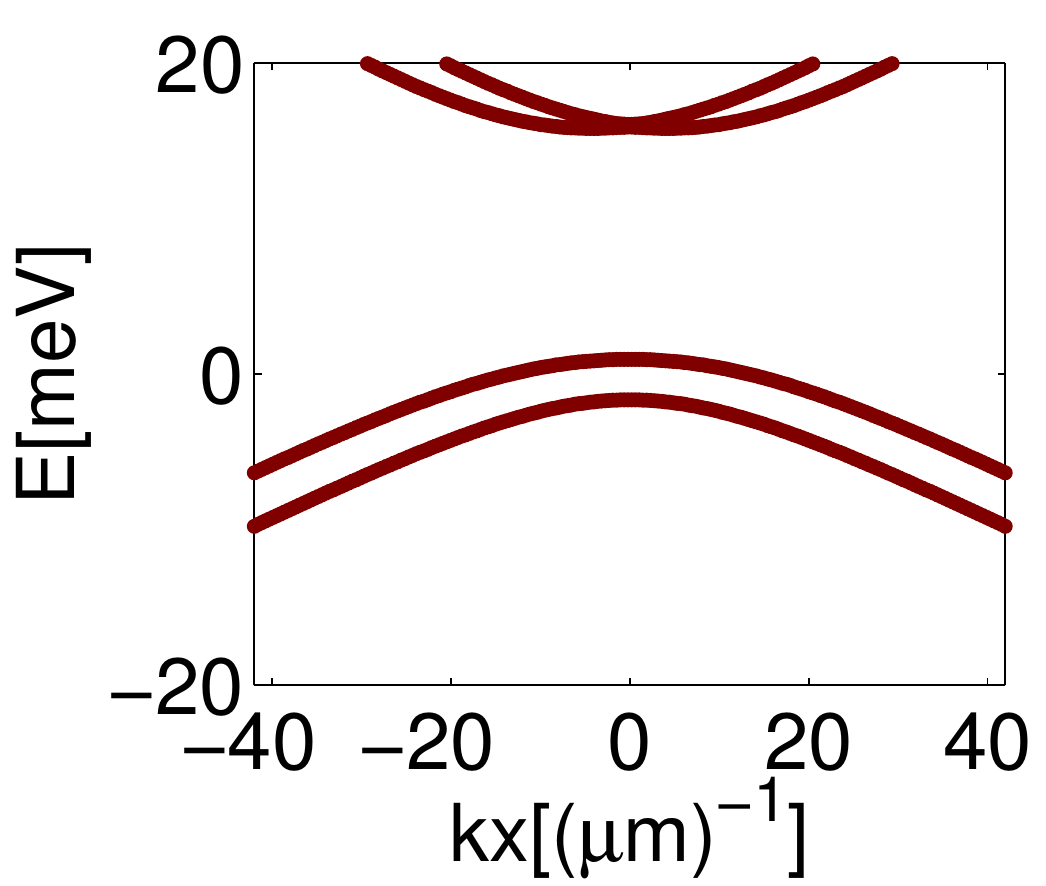}}
\subfigure[]{\includegraphics[width=0.3\textwidth]{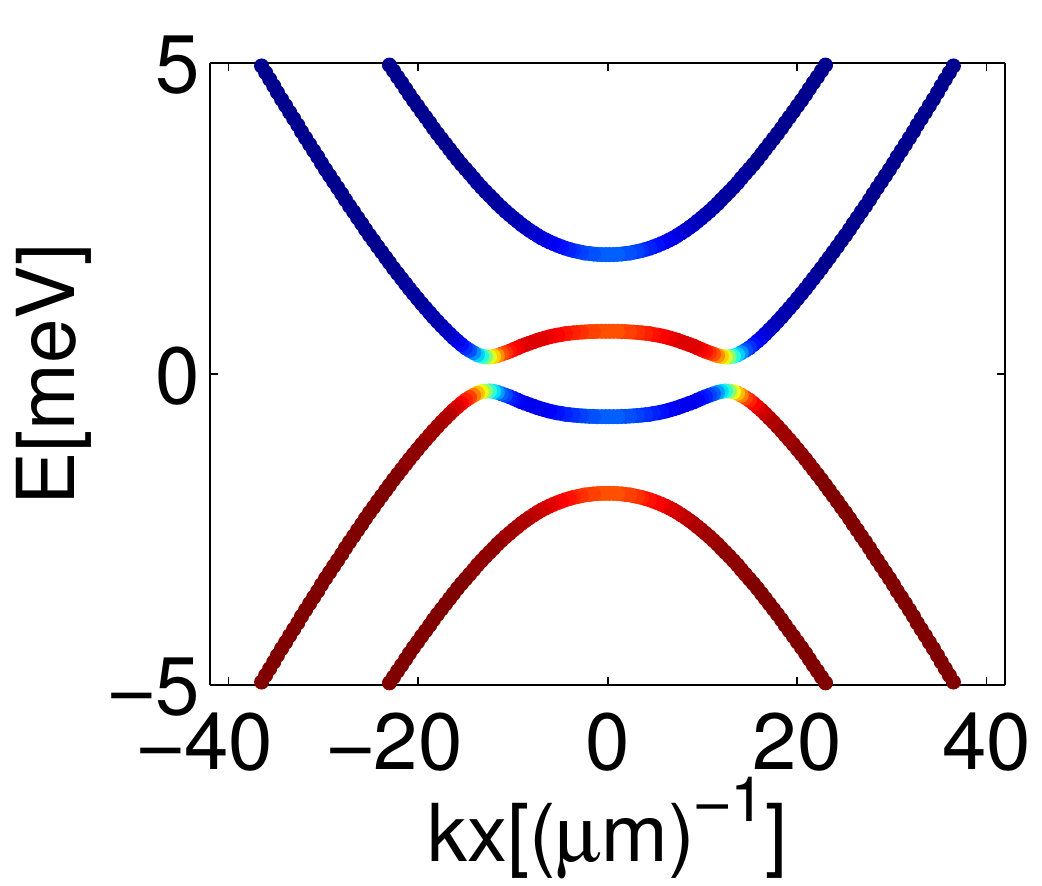}}
\end{center}
\caption{Left: Bulk energy dispersion of a HgTe-QW with doping
  parameter $C=7.8$\,meV in the presence of Zeeman terms and BIA-SOC; Rashba-SOC is absent and only electron states are shown. Right: The same parameters are
  used but the s-wave superconducting proximity terms are included. % with $\Delta_E=\Delta_H=0.5$\,meV. 
  The colors indicate the charge expectation value
  that ranges between e(red) and -e(blue).}
\label{fig:Ekbulk}
\end{figure}

Firstly, we show that BIA-SOC is sufficient in order to overcome the `fermion doubling problem'~\cite{Alicea2012} by freezing out half the degrees of freedom (unless $k=0$), which otherwise would make the spectrum doubly
degenerate. Under periodic boundary conditions,
$k_x$ and $k_y$ are good quantum numbers and the energy dispersion for the
standard BHZ Hamiltonian supplemented with the BIA term can be calculated to 
\begin{equation}
E(k_x,k_y)=\epsilon(k^2)\pm\sqrt{(A|k|\pm h)^2+ M(k^2)^2}.
\end{equation}
Clearly, the spectrum is non-degenerate for \mbox{$|k|\neq 0$}. By
additionally introducing the Zeeman terms, the time reversal symmetry is
broken and thus the Kramer's degeneracy at $|k|=0$ is lifted. \\
The combined effect of BIA and Zeeman terms on the bulk energy dispersion is
visualized in~\fref{fig:Ekbulk} (left). In the following, we choose the
Fermi energy $E_F$ to be zero so that the Fermi level is situated within the gap
between the $E1$-like bands opened by the Zeeman splitting. The doping
parameter is tuned in such a way that
$C\approx-M$. As a result, only a single
pair of Fermi points exists and thus the system has effectively spinless dynamics.  \par
When, in addition, the proximity of the s-wave superconductor is taken into
account, electrons with energy around the Fermi level and momenta $k$
and $-k$ couple, opening a gap around the Fermi level as shown
in~\fref{fig:Ekbulk} (right). The special conditions under which this gap closes
will be discussed in the following. The Fermi points at which this
happens are the key to the emergence of the chiral Majorana edge states~\cite{Volovik2003}. \par

\subsection*{Symmetry properties}
%Jetzt wird der volle Hamiltonian beschrieben, und zwar erst mal qualitativ nur die Symmetrie
Before calculating the Fermi points we will briefly review the symmetry
properties of the two $E1$-like bands closest to the Fermi
level. Their bulk energy spectrum is presented in~\fref{fig:Bulkspectrum2} in the presence of different
SOC terms: BIA or Rashba or the cooperation of BIA and
Rashba. Interestingly, although the spectrum is rotationally symmetric when either Rashba or BIA are
present, this rotational symmetry is broken when Rashba and BIA are
simultaneously present. \par
This can be understood as follows~\cite{Winkler2003}: 
While the effective magnetic fields due to Rashba-SOC are
oriented clockwise perpendicular to the wave vector $k$, the effective
magnetic fields due to BIA spin splitting have a tetrahedral symmetry, as
shown in~\fref{fig:Bulkspectrum2}. When both BIA and Rashba are present, the effective magnetic
fields add vectorially, decreasing or enhancing each other. The energy of the
lowest-energy band is very sensitive to the effective $k$-dependent in-plane
magnetic field. Therefore, it exhibits a tetrahedral symmetry when
BIA and Rashba interfere with each other. \par 

\subsection*{Fermi points}
In the following, we will identify the Fermi points of
the spectrum. To simplify analytical calculations, we project the full
Hamiltonian~\eref{eq:fullHamiltonian} onto an effective Hamiltonian. 
When $M$ defines the dominant energy scale and the doping is chosen as
$C\approx-M$, four of the eight bands become irrelevant.
Using quasi-degenerate perturbation theory~\cite{Winkler2003} (``L\"owdin
partitioning''), the eight-band Hamiltonian~\eref{eq:fullHamiltonian} can be reduced to
an effective four-band Hamiltonian acting on the $E1$-like electron and hole states only (see Appendix),
\begin{equation}
\eqalign{
\fl \tilde{H}=\mu(\hat{k}^2)\tau_z+B_E\sigma_z+\Delta_E\tau_x+R_0\hat{k}_x\sigma_y\tau_z-R_0\hat{k}_y\sigma_x\tau_z\cr
+\tilde{h}\hat{k}_x\left(\cos{2\theta}\sigma_x-\sin{2\theta}\sigma_y\right)\tau_z-\tilde{h}\hat{k}_y\left(\cos{2\theta}\sigma_y+\sin{2\theta}\sigma_x\right)\tau_z.
}
\label{eq:effectivehamiltonian}
\end{equation}
where the new parameters $\mu(k^2)=\epsilon(k^2)+M(k^2) +(A^2k^2+h^2)/(2M)$
and $\tilde{h}=Ah/M$ have been introduced. \par
In the absence of BIA-SOC, the reduced form~\eref{eq:effectivehamiltonian} is analogous
to the effective Hamiltonians of different semiconductor systems, which have
already been shown to host chiral Majorana edge
states~\cite{Sau2010, Ghosh2010, Oreg2010}. Note that the Rashba energy $U_{R}\equiv
2m^{*}R_{0}^2$~\cite{Potter2011} is kept quite small here by intention in order to study the combined effect of
BIA and Rashba-SOC: for the
values in table 1 it amounts to \mbox{$U_{R}=R_{0}^{2}/(\frac{A^2}{2M}-D-B) \approx
0.3$\,meV}. On the other hand, the BIA-SOC leads to an effective
  Dresselhaus-SOC in the reduced model \eref{eq:effectivehamiltonian}, which amounts to $U_{D}=(Ah/M)^{2}/(\frac{A^2}{2M}-D-B) \approx
0.7$\,meV. In~\cite{Virtanen2012}, electric fields up to $\mathcal{E}\sim
100$\,mV/nm were estimated to be possible, which could enhance the Rashba
energy in principle by a factor of 30, {without affecting the effective Dresselhaus energy.\par
The effective Hamiltonian~\eref{eq:effectivehamiltonian} yields (for $\theta=\pi/4$ and
$\Delta_E\neq 0$) zero energy level crossings at
\begin{equation}
(R_0-\tilde{h})^2k_x^2+(R_0+\tilde{h})^2k_y^2=-\left(|\Delta_E|\pm\sqrt{B_E^2-\mu(k^2)^2}\right)^2.
\label{eq:topologicaltransition}
\end{equation}
This implies that for
non-vanishing SOC, Fermi points exist (for $B_E^2-\mu(k^2)^2>0$) at 
\begin{eqnarray}
k_x=0, k_y=0 \textrm{ and } |\Delta_E|=\sqrt{B_E^2-\mu_0^2}\label{eq:fermi1}\\
k_x=0, k_y\neq 0, R_0=-\tilde{h} \textrm{ and~\eref{eq:topologicaltransition}}\label{eq:fermi2}\\
k_x\neq 0, k_y=0, R_0=\tilde{h} \textrm{ and~\eref{eq:topologicaltransition}}\label{eq:fermi3}
%k\neq 0, R_0=\tilde{h}=0 \textrm{ and~\eref{eq:topologicaltransition}}.\label{eq:fermisurface}
\end{eqnarray}
where $\mu_0=\mu(k^2=0)=C+M+h^2/(2M)$. %More precisely,~\eref{eq:fermi1},~\eref{eq:fermi2},~\eref{eq:fermi3} define isolated Fermi
While the Fermi point~\eref{eq:fermi1} %and the Fermi surface 
exists also in other semiconductor systems that have been proposed to host
chiral Majorana edge
states~\cite{Sau2010, Ghosh2010, Oreg2010}, the Fermi points~\eref{eq:fermi2} and~\eref{eq:fermi3} are characteristic of a system with two different SOC terms.

\begin{figure*}[hbtp]
\begin{tabular}{ccc}
\subfigure[Rashba]{
\includegraphics[width=0.30\textwidth]{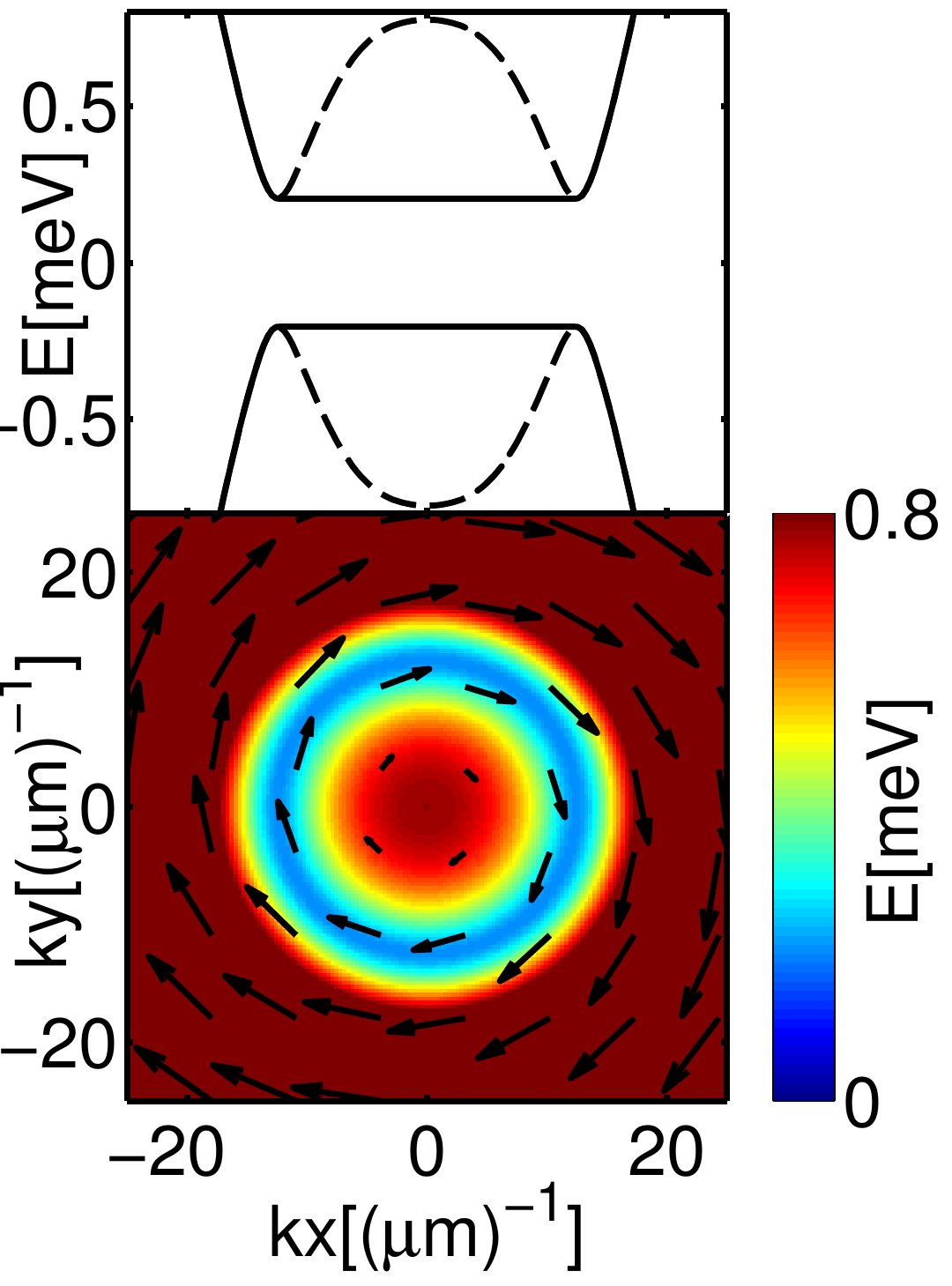}
}&
\subfigure[BIA]{
\includegraphics[width=0.30\textwidth]{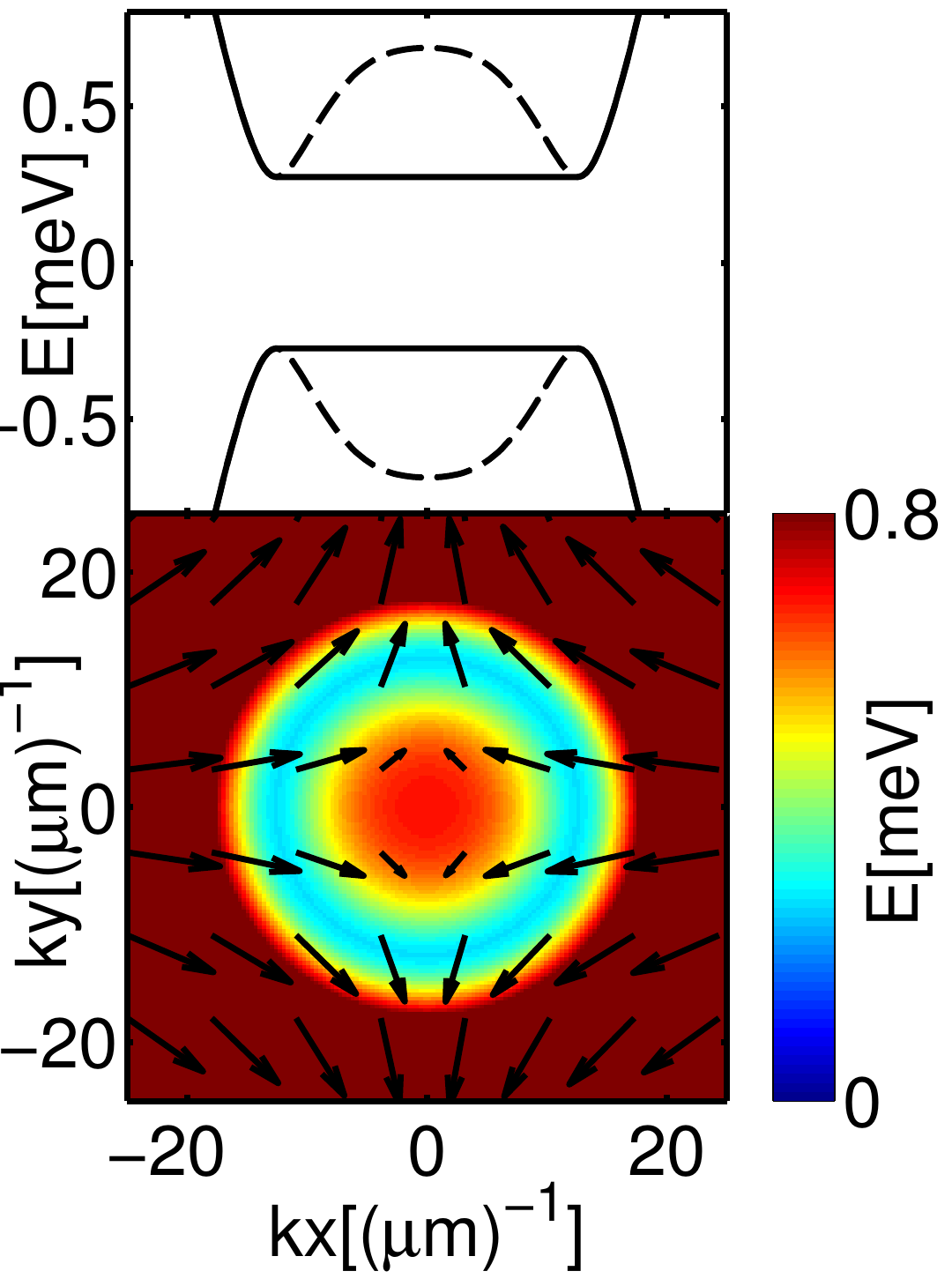}
}&
\subfigure[BIA and Rashba]{
\includegraphics[width=0.30\textwidth]{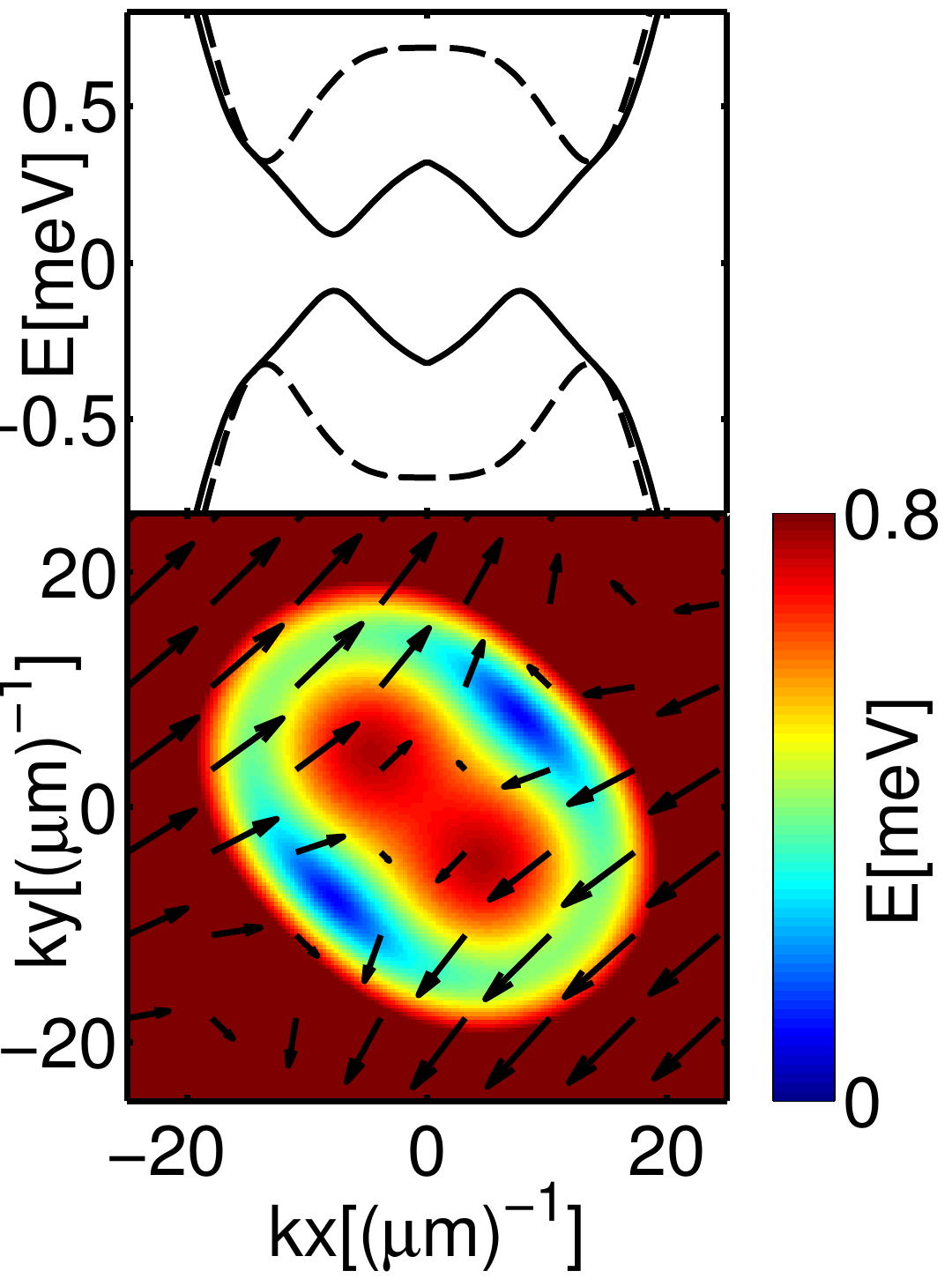}
}
\end{tabular}
\caption{Bulk energy dispersion when a) only the Rashba term is included
  b) only the BIA term is included c) both terms are included. Top: energy
  dispersion $E(k_x)$ where $k_y=0$ (dashed line) compared to energy dispersion
  $E_{min}(k_x)$ where $k_y$ is such that the energy becomes minimal (solid line). Bottom:
  energy dispersion $E(k_x,k_y)$ for the minimal positive energy band. Vectors
  illustrate in-plane spin expectation
  values projected onto the E1-band $\langle \sigma_x(1+s_z)/2 \rangle$ and $\langle  \sigma_y(1+s_z)/2 \rangle$.
}
\label{fig:Bulkspectrum2}
\end{figure*}

\subsection*{Bulk Topological Invariant}
In this section, we identify the topological phases of the non-trivial
momentum space topology of Hamiltonian~\eref{eq:fullHamiltonian}.  
The fundamental Hamiltonian~\eref{eq:fullHamiltonian} belongs to
the Cartan-Altland-Zirnbauer symmetry class $D$, which in two
spatial dimensions is characterized by a topological invariant $\in
\mathds{Z}$~\cite{Altland1997}. The first Chern number which is commonly used
as the topological invariant can be obtained for a
non-interacting Hamiltonian with non-degenerate spectrum as the integral of the Berry curvature over the first Brillouin zone 
\begin{equation}
\mathcal{C}_1=\frac{1}{2\pi}\int_{\mathrm{BZ}} \mathcal{F}\rmd^2k\\
\label{eq:C1}
\end{equation}
where the Berry curvature
$\mathcal{F}=\partial_{k_x}\mathcal{A}_{k_y}-\partial_{k_y}\mathcal{A}_{k_x}$ is obtained from the
the Bloch functions via $\mathcal{A}_{k_\alpha}=-i\sum_j\langle
u_j(k)|\partial_{k_\alpha}|u_j(k)\rangle$, $\alpha=x,y$, the index $j$ running over the occupied states.\par
For non-vanishing spin-orbit interactions~\footnote{i.\,e. BIA and Rashba should not both be zero}, the first Chern number of the
Hamiltonian~\eref{eq:fullHamiltonian} is given by 
\begin{equation}
\mathcal{C}_1=\Theta{\left(B_E^2-\mu_0^2-\Delta_E^2 \right)} \sign{\left(R_0^2-\tilde{h}^2\right)},
\label{eq:C1eval}
\end{equation}
where $\Theta(x)$ is the Heaviside step function. This was verified by computing the
topological invariant from a discretized version~\cite{Resta2012} of~\eref{eq:C1} for a regularized
version\footnote{i.e. replacing $k_x^2+k_y^2\rightarrow
  \frac{2}{a^2}\left(2-\cos(ak_x)-\cos(ak_y)\right)$, $k_x\rightarrow
  \frac{1}{a}\sin(ak_x)$, $k_y\rightarrow   \frac{1}{a}\sin(ak_y)$ where
  $a$ is a lattice constant that is chosen in such a way that no artificial
  level crossings occur.} of Hamiltonian~\eref{eq:fullHamiltonian} numerically using MATLAB at several illustrative
values of the parameters on each side of the topological
transitions. Equation~\eref{eq:C1eval} is valid when the BIA parameter $h$ is
  small compared to the mass parameter $M$. \par
While the general form of~\eref{eq:C1eval} is analogous to
known topological invariants of similar systems such as spin-obit coupled quantum
wires~\cite{Oreg2010} or 2D spin-orbit coupled semiconductors~\cite{Sau2010}, this constitutes to our knowledge the first such
calculation for a HgTe-QW in proximity to a
superconductor. Importantly, this is also the first time that the BIA term $h$
is included. When both Rashba and BIA are present, the sign of the
topological invariant $\mathcal{C}_1$ can be reversed due to the interplay between both
SOCs. This is similar to the combined effect of Rashba and
Dresselhaus SOCs that was described for non-centrosymmetric superconductors~\cite{You2012}. \par     
In the absence of BIA, the relevant parameters for a topological transition to occur are the proximity coupling $\Delta_E$, the
Zeeman energy $B_E$ and the sum of mass gap and doping, $\mu_0=C+M$. When BIA is taken into account, the topological transition is shifted towards higher doping as described by the term $\mu_0=C+M+\frac{h^2}{2M}$ (in the
limit that $M$ is the dominant energy scale). The topological transition is presented as a function of Zeeman energy
$B_E$ and doping C in~\fref{fig:topologicaltransition}.
\begin{figure}[hbtp]
\centering
\includegraphics[width=0.5\textwidth]{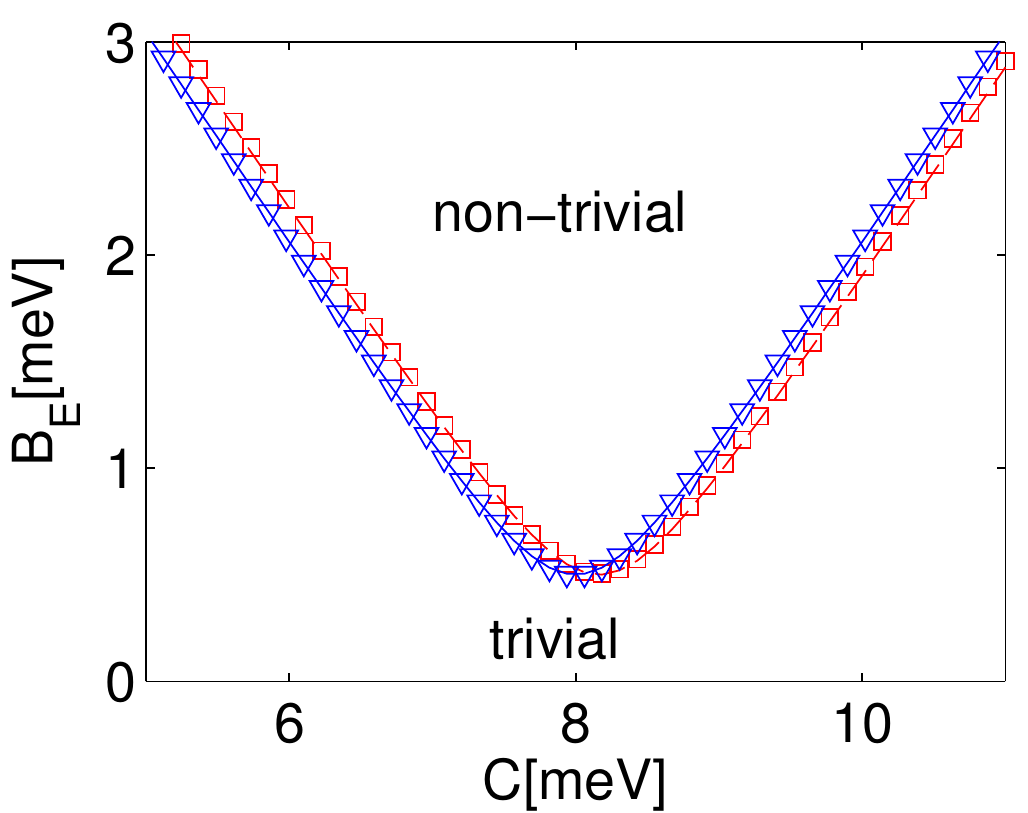}
\caption{Topological phase diagram in the parameter region Zeeman energy
  $B_E$ and doping C. The topological transition described approximatively
  by~\eref{eq:C1eval} (blue solid line when BIA is present, red dashed line when BIA
  is absent) is compared to numerical calculations based
  on~\eref{eq:fullHamiltonian} (blue triangle when BIA is present, red square
  when BIA is absent). When BIA asymmetry is taken into account, the topological transition is shifted
towards higher doping. The other parameters are given in~\tref{tab:HgTeValues}.
}
\label{fig:topologicaltransition}
\end{figure}

\section{Majorana Edge States}\label{sec:Edge} 
In this section, we discuss the appearance of Majorana edge states in this system. \\
%As described in \cite{Oreg2010} (but for a quantum wire), the phase transition which is evident in 
%\[E_0=|E_Z|-\sqrt{\Delta^2+\mu^2}\]
%allows Majorana fermion states to form. Here, $-\mu=C+M+\frac{h^2}{2M}+\mathcal{O}(|k|^2)$
%abbreviates properties of the quantum well.
With an edge we mean that there is a domain wall in the shape of a straight edge
where one side is topologically nontrivial ($\mathcal{C}_1=\pm 1$) and the other side is
topologically trivial ($\mathcal{C}_1=0$). 
We distinguish between two cases:
either (i) the topologically nontrivial region is vacuum or (ii) a topologically
trivial QW. In the following, these cases will be referred as (i) 
hard-wall boundary conditions and (ii) soft-wall boundary
conditions. Soft-wall boundary conditions (two different topological
phases in a QW) can be achieved by varying the
doping parameter $C$ accordingly across the sample, as shown
in~\fref{fig:WaveFunctions} (b). \par
%In both cases, the value of the invariant changes at the boundary, implying
%that the energy gap closes at the interface, leading to edge states. 
%The edge states will be explicitely calculated in the following. To this end, we . As
%stated previously, this coincides with the [100] crystal
%direction for $\theta=0$. 
%\footnote{The calculation is equivalent for an edge in [010]
%  direction. In an arbitrary in-plane crystal direction however, the results
%  would be considerably different. We assume that cleavage along the crystal
%  direction is experimentally most accessible.}. 
%\par
Without loss of generality, we assume that the domain wall
lies parallel to the $x$-axis and that the region $y<0$ is topologically
non-trivial while $y>0$ is topologically trivial. 
Due to translational invariance along the $x$-axis we can consider solutions
that propagate along the edge with definite $k_x$. 
\par
The general strategy when solving such a boundary problem is as follows:
In a first step, we use the Ansatz
\begin{equation}
\psi(x,y,E)=e^{ik_xx}e^{ik_yy}\psi(k_x,k_y),
\end{equation}
to calculate the wave functions on both sides of the boundary. Hereby, $k_y$,
which is not a good quantum number, has to be determined as a function of $k_x$
and the energy $E$ separately for both sides of the boundary. In addition, the
eigenvectors are determined as a function of $k_x$
and $E$. \\
In the second step, the boundary conditions are applied, finally yielding an
energy dispersion $E(k_x)$. For hard wall boundary conditions, the wave function is required
to vanish at the boundary between the HgTe sample and vacuum. For soft wall
boundary conditions, both the wave function and its first derivative are required
to be continuous at the boundary between the topologically trivial and the
topologically non-trivial HgTe samples. In both cases, we require that the solutions are bound to the edge, i.e. that
the probability densities vanish far away from the boundary (see~\fref{fig:WaveFunctions}). 
\begin{figure}[htbp]
\centering
\subfigure[]{\includegraphics[width=0.4\textwidth]{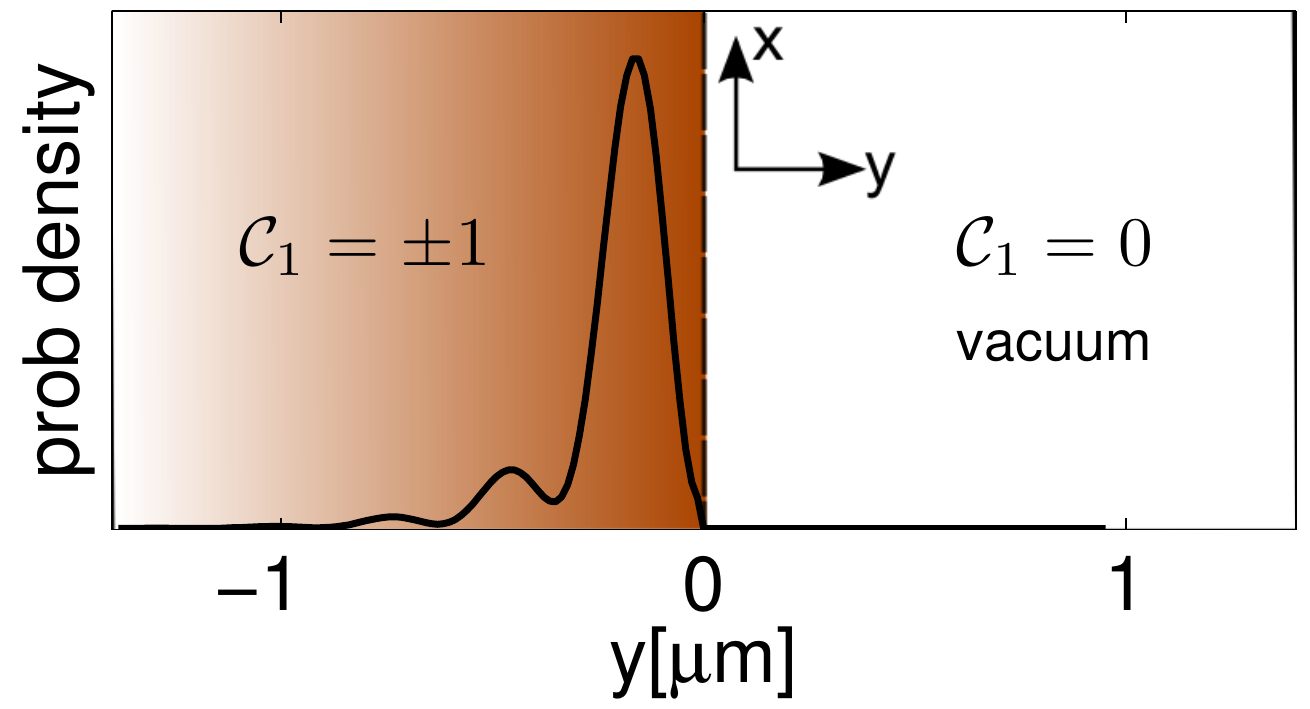}}
\subfigure[]{\includegraphics[width=0.4\textwidth]{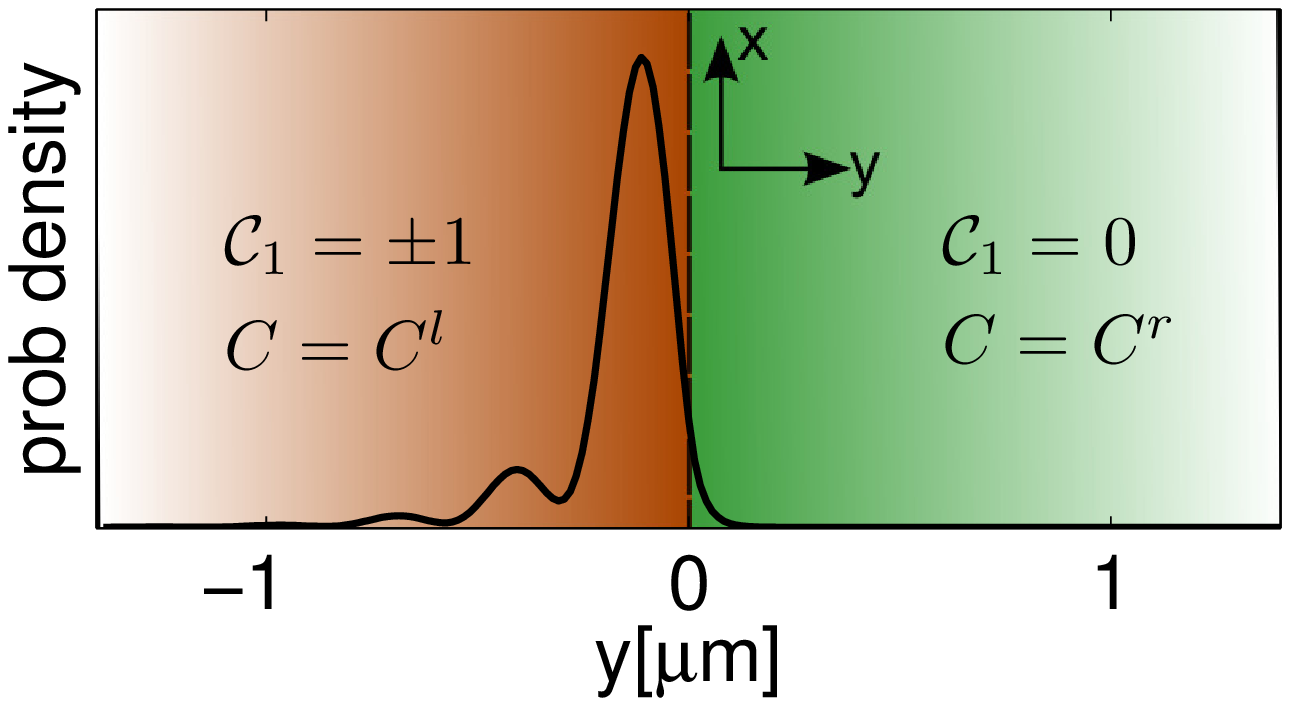}}
\caption{Normalized probability density at $k_x=0$ as a function of $y$ for soft wall
  boundary conditions separating a topologically non-trivial QW from vacuum (trivial)
  (a) and hard wall boundary conditions separating a topologically non-trivial
  QW region from a topologically non-trivial QW region.}
\label{fig:WaveFunctions}
\end{figure}

\begin{figure*}[htbp]
\centering
\subfigure[]{\includegraphics[width=0.3\textwidth]{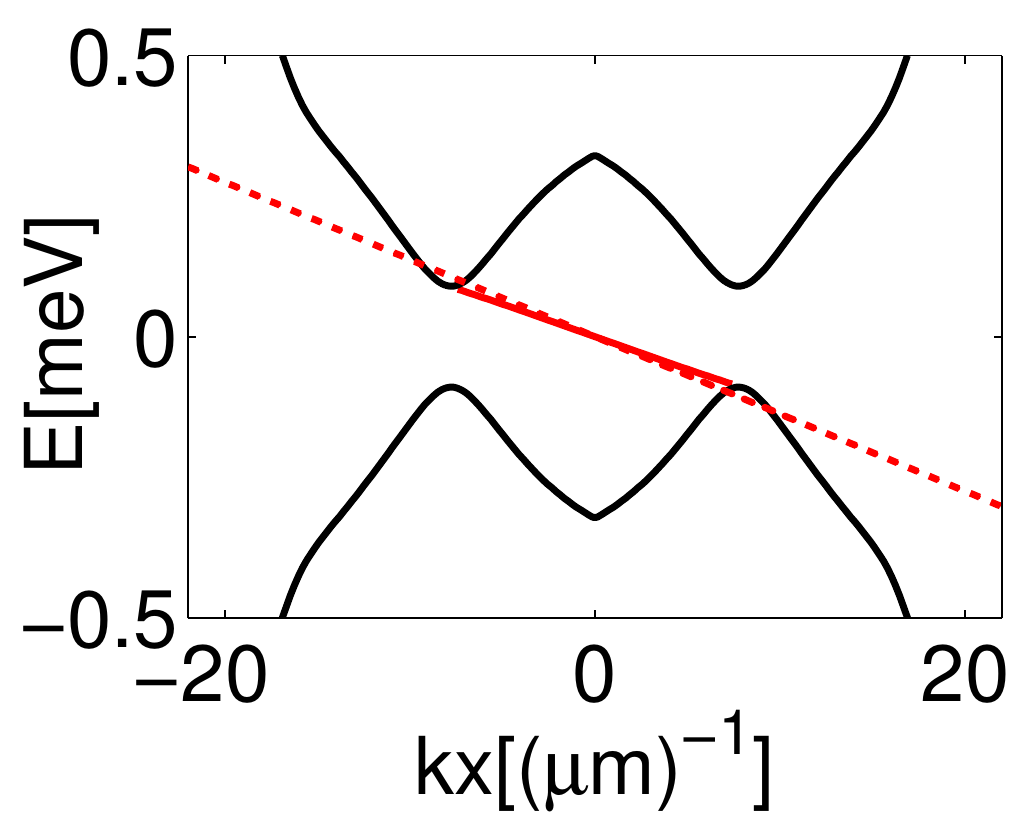}}
\subfigure[]{\includegraphics[width=0.3\textwidth]{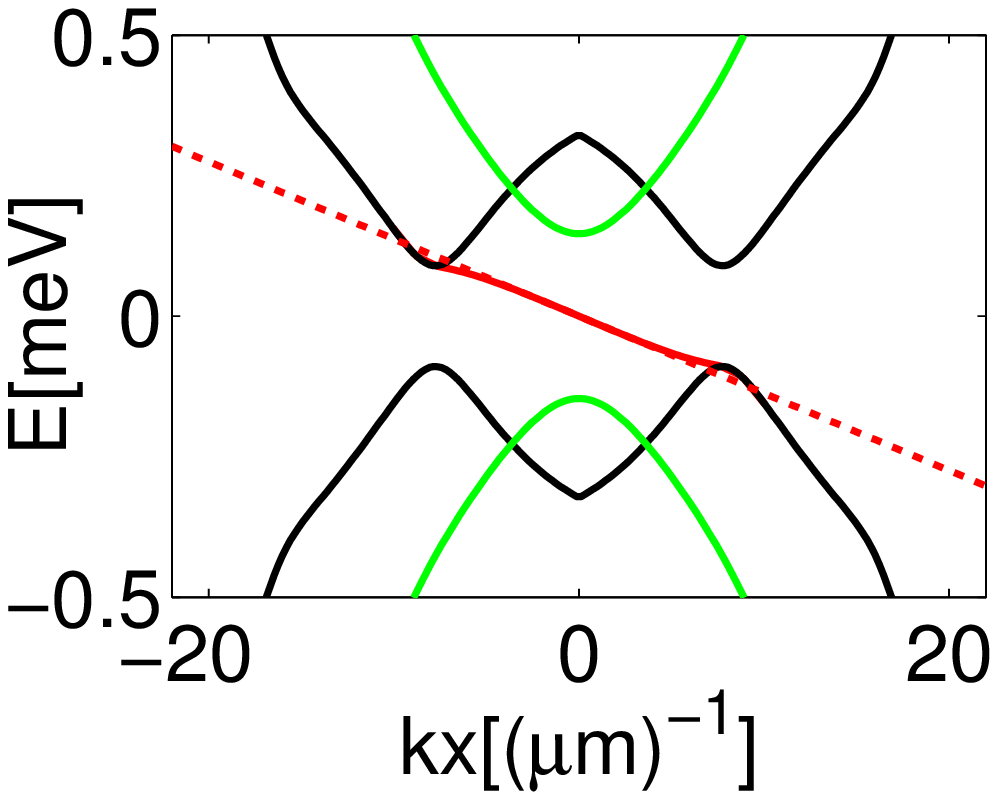}}
\caption{Energy dispersion with a boundary in $x$-direction. On the left hand side of
  the boundary there is a topologically non-trivial QW with $C^l$=7.8\,meV
  (black bulk energy minimum), while on the right
hand side of the boundary there is either vacuum (a) or a topologically
trivial QW with $C^r$=6.8\,meV, indicated by the green bulk energy minimum (b). Chiral
  Majorana edge states obtained numerically as boundary solutions to~\eref{eq:fullHamiltonian} are shown in red. The red
dotted lines are estimations derived from~\eref{eq:MajoranaVelocity}. The
parameter values are taken from~\tref{tab:HgTeValues}.}
\label{fig:single_boundary}
\end{figure*}

\subsection*{Calculation of edge state dispersion for small $k_x$}
Under certain simplifying assumptions, the edge dispersion of the boundary
problem can be calculated to first order in $k$ using perturbation theory. 
We assume that is is sufficient to calculate the edge dispersion in the
reduced model~\eref{eq:effectivehamiltonian} and require that quadratic terms
$\propto k_x^2, k_y^2$ may be neglected (Assumption 1). Consistently, we also match only the wave functions and not the
derivatives at the boundary.\\
We first prove that the boundary conditions are fulfilled for $E=0, k_x=0$ and
calculate the wave function of the corresponding bound state. \par
Imposing the condition $E=k_x=0$ in~\eref{eq:effectivehamiltonian} yields four possible values for $\lambda=ik_y$:
\begin{equation}
\pm\lambda_\pm=\frac{\pm\left( |\Delta_E|
  \pm\sqrt{B_E^2-\mu_0^2}\right)}{\sqrt{R_0^2+\tilde{h}^2+2R_0\tilde{h}\sin{2\theta}}} .\\
\end{equation}
Depending on the parameter values, the signs of $\lambda$ can be determined as
follows under the assumption that $|B_E|>|\mu_0|$:
\begin{center}
\begin{tabular}{c|c}
$B_E^2>\mu_0^2+\Delta_E^2$ &
$B_E^2<\mu_0^2+\Delta_E^2$ \\
\hline
$ +\lambda_+ >0$ &  $+\lambda_+>0$ \\
$ -\lambda_- >0$ &  $+\lambda_->0$ \\
\end{tabular}
\end{center}
The respective eigenvectors are 
\begin{equation}
|\lambda_+\rangle =  \frac{1}{2}\left(
\begin{array}{c}
e^{i\chi}\sqrt{1-\mu_0/B_E}\\ 
-\sign{(B_E)}\sqrt{1+\mu_0/B_E}\\
e^{i\chi}\sign{(B_E)}\sqrt{1+\mu_0/B_E}\\
\sqrt{1-\mu_0/B_E}
\end{array}
\right)  
%|k_{y,-}\rangle &=  \frac{1}{2B}\left(e^{i\frac{\chi}{2}}\sqrt{B+\mu},  e^{-i\frac{\chi}{2}}\sqrt{B-\mu},-e^{i\frac{\chi}{2}}\sqrt{B-\mu}, e^{-i\frac{\chi}{2}}\sqrt{B+\mu}\right)^T\\ 
%|-k_{y,+}\rangle &= \frac{1}{2B}\left(e^{i\frac{\chi}{2}}\sqrt{B+\mu},  e^{-i\frac{\chi}{2}}\sqrt{B-\mu},e^{i\frac{\chi}{2}}\sqrt{B-\mu}, -e^{-i\frac{\chi}{2}}\sqrt{B+\mu}\right)^T\\  
%|-k_{y,-}\rangle &=\frac{1}{2B}\left(-e^{i\frac{\chi}{2}}\sqrt{B+\mu},  e^{-i\frac{\chi}{2}}\sqrt{B-\mu},e^{i\frac{\chi}{2}}\sqrt{B-\mu}, e^{-i\frac{\chi}{2}}\sqrt{B+\mu}\right)^T
\end{equation}
with $e^{i\chi}=\left(iR_0+\tilde{h}e^{2\theta i}\right)/\sqrt{R_0^2+\tilde{h}^2+2R_0\tilde{h}\sin{2\theta}}$
and where
\mbox{$|\lambda_-\rangle$}, \mbox{$|-\lambda_+\rangle$} and
\mbox{$|-\lambda_-\rangle$} can be obtained from
\mbox{$|\lambda_+\rangle$} by clockwise translation of the minus sign
in front of the second entry. \par
If $B_E^2>\mu_l^2+\Delta_E^2$ on the left side of the
boundary (topologically non-trivial) and $B_E^2<\mu_r^2+\Delta_E^2$ on the right side
of the boundary (topologically trivial), the matching condition requires that
the eigenvectors on the left and right hand sides of the boundary are linearly dependent, 
\begin{equation}
\det{\left(|\lambda^l_+\rangle , |-\lambda^l_-\rangle,|-\lambda^r_+\rangle,|-\lambda^r_-\rangle\right)}   =0
\end{equation}
yielding the solution 
\begin{equation}\fl
\psi(y)=\frac{1}{\mathcal{N}}\left(\Theta{(-y)}
e^{-\lambda^l_-y}|-\lambda^l_-\rangle +\Theta{(y)}\left(c_1 e^{- \lambda^r_+y}|-\lambda^r_+\rangle + c_2e^{- \lambda^r_-y}|-\lambda^r_-\rangle\right)\right) .
\end{equation}
Here $\mathcal{N}$ is a normalization factor such that $\int \rmd y
|\psi(y)|^2=1$. The coefficients $c_1$ and $c_2$ are determined by the matching
condition.\par 
This proves that there is a bound state for \mbox{$k_x=E=0$}. The antisymmetry of the Hamiltonian~\eref{eq:fullHamiltonian} under
particle-hole transformation $\Xi^\dagger H\Xi = -H$, where $\Xi=iK\sigma_y\tau_y$, has the important consequence that $\psi(y)$ for \mbox{$k_x=E=0$} constitutes
a Majorana excitation, as can be easily confirmed by explicit calculation. \par
%it
%constitutes an eigenstate of the particle hole operator $\Xi$, 
%\[\Xi\psi(y)=\psi(y).\] \par
Now we consider the case of non-zero $k_x$. When $k_x$ is small, the
$k_x$-dependent BIA and
Rashba terms in equation~\eref{eq:effectivehamiltonian} can be treated as a
perturbation $V$. The expectation value of this perturbation with respect to
the $k_x=E=0$ state gives the energy in terms
of $k_x$ to first order in perturbation theory,   
\begin{equation}
E=\int \rmd y \psi^*(y)V\psi(y)=\frac{|\Delta_E|}{B_E}\frac{R_0^2-\frac{A^2h^2}{M^2}}{\sqrt{R_0^2+\frac{A^2h^2}{M^2}+2R_0\frac{Ah}{M}\sin{2\theta}}}k_x.
\label{eq:MajoranaVelocity}
\end{equation}

\subsection*{Edge state dispersions}
The resulting edge energy dispersions that were calculated numerically from~\eref{eq:fullHamiltonian} and
using~\eref{eq:MajoranaVelocity} are presented and compared to the bulk
energy dispersions in~\fref{fig:single_boundary}. \par
The boundary between topologically trivial and non-trivial regions results in
a single chiral Majorana bound state of
approximately linear energy dispersion that smoothly connects to the occupied
and non-occupied bulk bands. The existence of this edge state
is independent of which boundary condition (soft or hard wall) or which
inversion asymmetry term (BIA, Rashba or both) is considered. As expected, its
existence depends on the topological bulk properties on both sides of the
boundary - the edge state disappears when the Chern numbers on both sides of
the boundary are identical.\par 
The formula~\eref{eq:MajoranaVelocity} describes the numerically obtained energy dispersions
of~\fref{fig:single_boundary} well, regardless of whether hard wall or soft
wall boundary conditions have been employed. A small discrepancy between
the numerical calculation on the basis of the
Hamiltonian~\eref{eq:fullHamiltonian} and~\eref{eq:MajoranaVelocity} is mostly due
to the limited applicability of Assumption 1. Yet,~\eref{eq:MajoranaVelocity}
describes the salient features of the edge dispersion well. \par
As predicted by~\eref{eq:MajoranaVelocity}, the edge state velocity depends
most strongly on the parameters $\theta, A, h, R_0, B_E, \Delta_E$. 
We have checked that the parameters $D, B, C, \Delta_H, B_H$ have comparably little effect on the energy dispersion.\par 
Interestingly, we find that the group velocity $v=\frac{1}{\hbar}\frac{\partial E}{\partial
  k_x}$ of the chiral edge state is tunable by the interplay between the Rashba and
BIA spin-orbit interactions. The edge state is right-moving when the Rashba
term dominates and left-moving when the BIA term dominates
(see~\fref{fig:single_boundary_properties4}). This is consistent with the change of the
first Chern number in~\eref{eq:C1eval} between
$\mathcal{C}_1=\pm 1$ by means of the index theorem~\cite{Volovik2003}. When applied to the
hard-wall boundary condition, the index theorem equates the
topological charge $\mathcal{C}_1$ of the bulk with the
sign $\nu$ of the slope of the chiral edge
mode. It is therefore easy to see that as the Rashba parameter $R_0$
changes from $R_0^2<\tilde{h}^2$ to $R_0^2>\tilde{h}^2$, the first Chern
number changes from $-1$ to $+1$ and the edge state from left to right
moving. This could be experimentally verified by changing the Rashba
parameter, while the intrinsic BIA-SOC remains fixed. Another interesting
observation is that the group velocity of the chiral Majorana edge state
depends on the orientation of the edge with respect to the crystal axis when
both Rashba and BIA SOC terms are present. The absolute values of the velocity
is smallest along the $[110]$ ($\theta=\pi/4$) and largest along the
$[\bar{1}10]$ ($\theta=3\pi/4$) direction. In HgTe-QWs, the edge structures
are for experimental reasons prepared in parallel to the cleavage planes, and
therefore 95 $\%$ of the current carrying edges are orientated either in
$[110]$ or $[\bar{1}10]$ direction~\cite{Buhmann}. Due to the simultaneous
presence of Rashba and BIA SOC terms, these two experimentally accessible edge
orientations are expected to have distinct chiral Majorana edge state properties.

\begin{figure*}[htbp]
\centering
\subfigure[]{\includegraphics[width=0.30\textwidth]{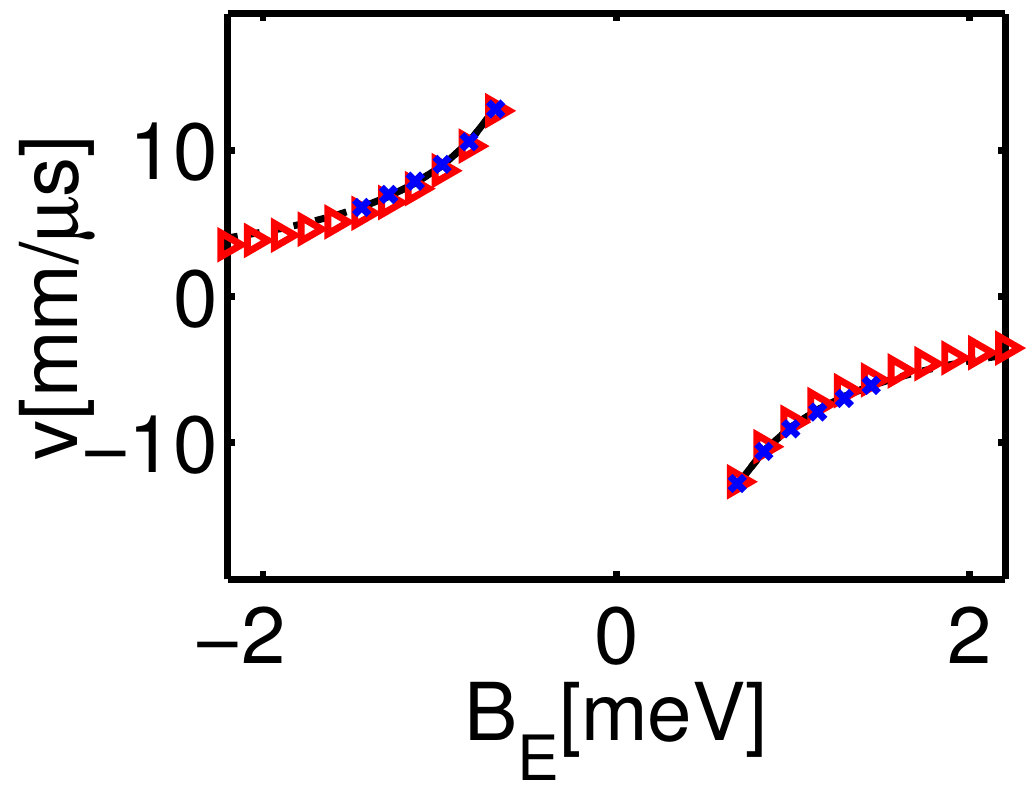}}
\subfigure[]{\includegraphics[width=0.30\textwidth]{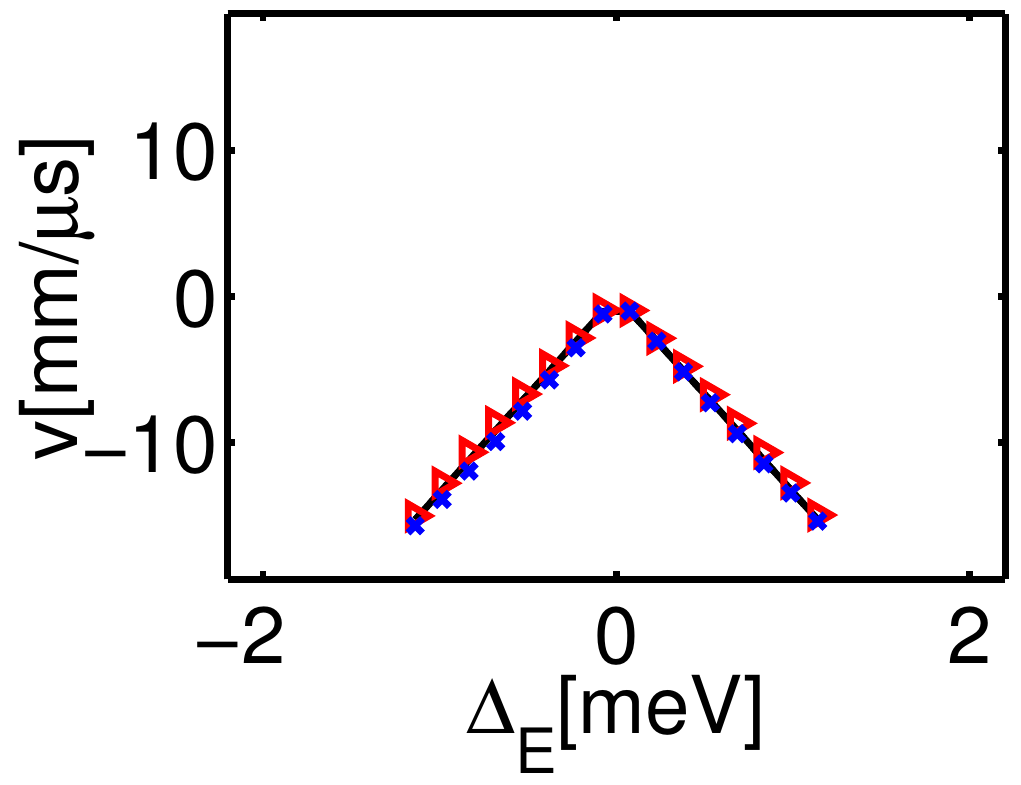}}
\subfigure[]{\includegraphics[width=0.30\textwidth]{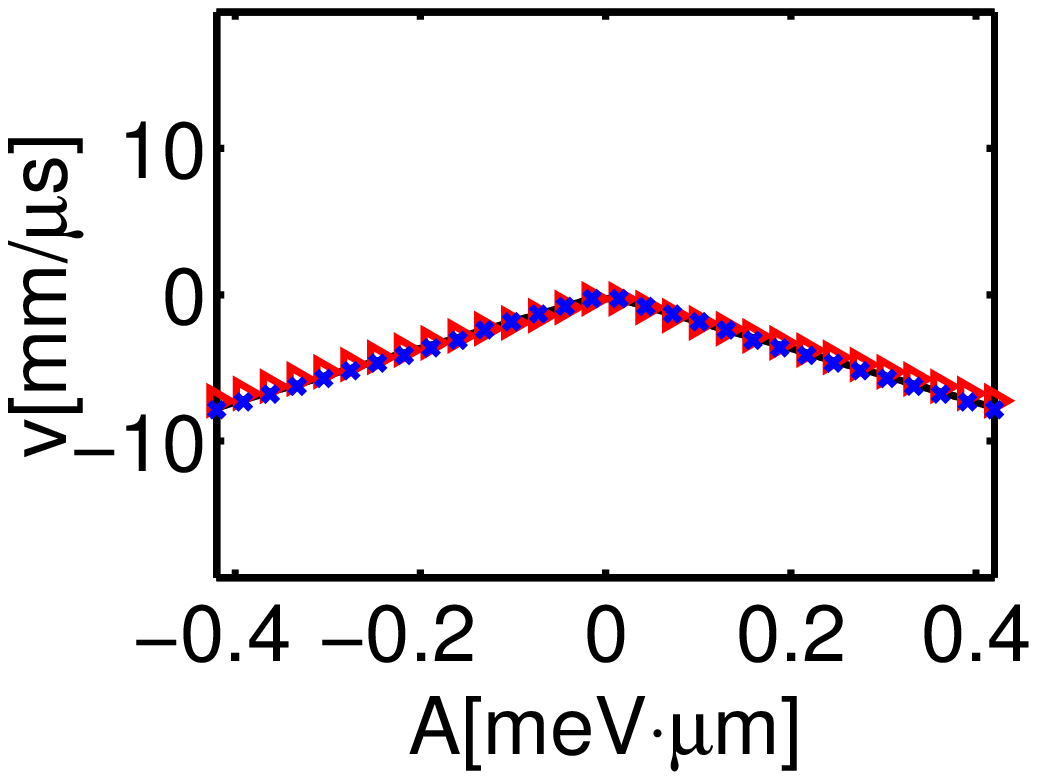}}\\
\subfigure[]{\includegraphics[width=0.30\textwidth]{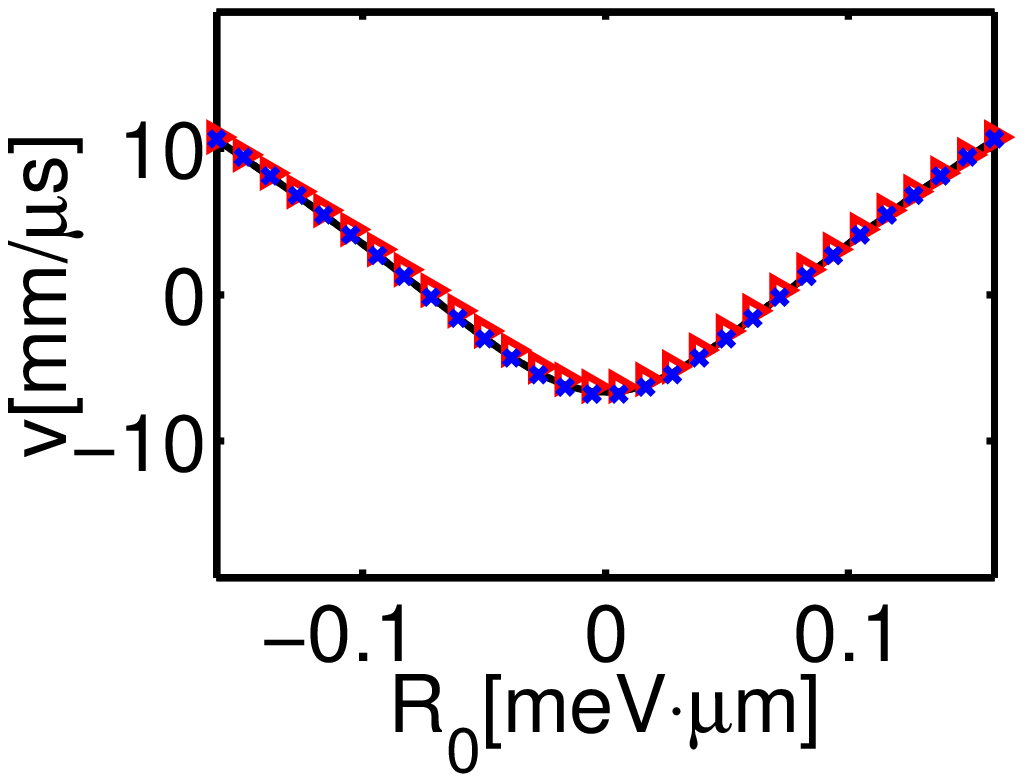}}
\subfigure[]{\includegraphics[width=0.30\textwidth]{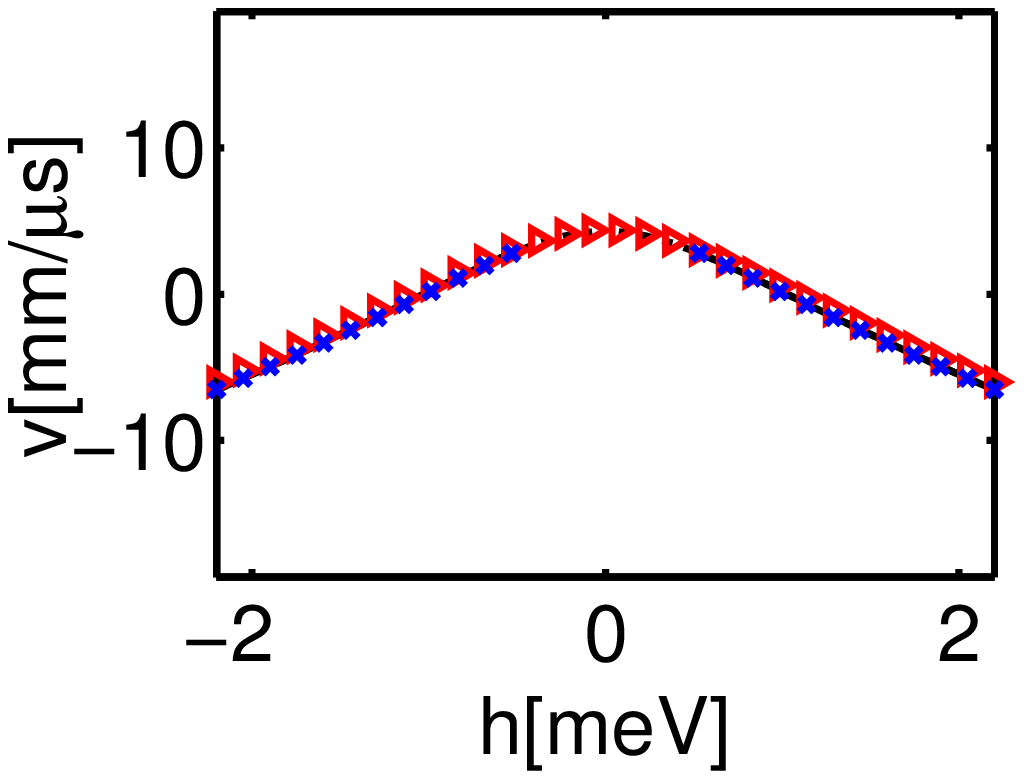}}
\subfigure[]{\includegraphics[width=0.30\textwidth]{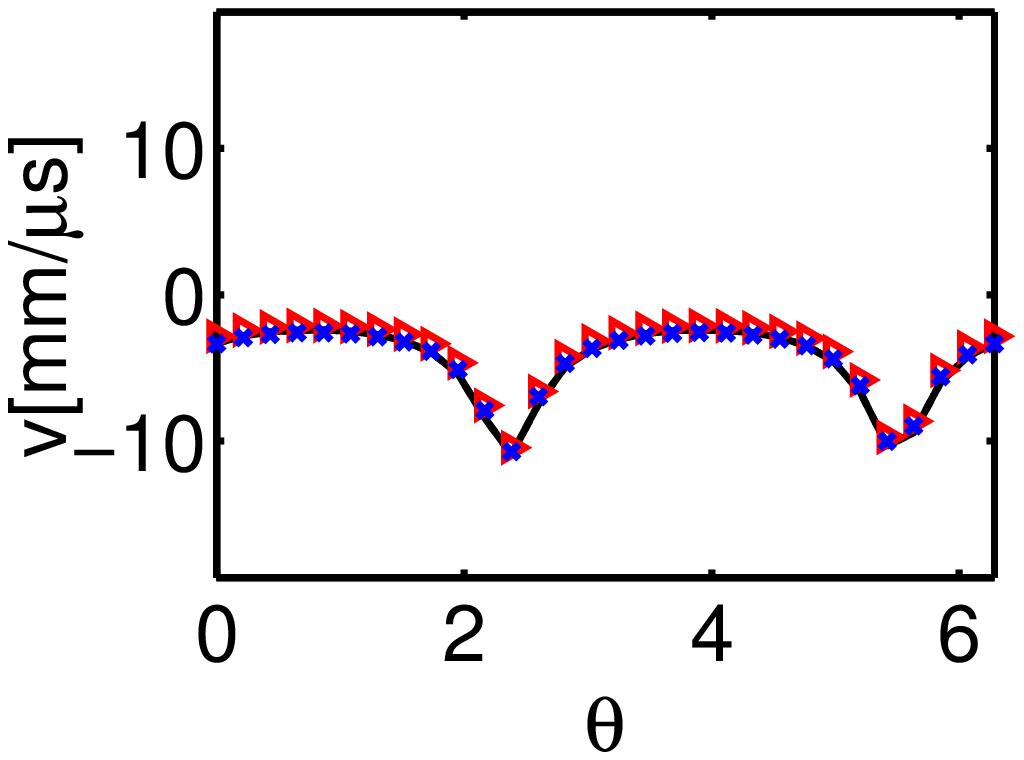}}
\caption{The edge group velocities $v=\frac{1}{\hbar}\frac{\partial E}{\partial k_x}$ as a function of different
  parameters. Analytic approximations (solid black lines) obtained
  from~\eref{eq:MajoranaVelocity} are compared with numerical values for hard
  wall (red triangles) and soft wall boundary conditions (blue crosses). This
  confirms that the edge state only exists when the boundary connects
  topologically trivial and non-trivial regions, compare~\eref{eq:C1eval}. 
The Rashba spin-orbit coupling is set to
zero except in the last line; all other parameters are taken from~\tref{tab:HgTeValues}.}
\label{fig:single_boundary_properties4}
\end{figure*}

\section{Comparison of chiral Majorana edge states and helical edge states}\label{sec:Comparison}
As described in the previous section, a HgTe-QW in proximity to a superconductor
with BIA and a Zeeman field hosts chiral Majorana edge states at its edges when the
parameters are in the topological nontrivial TSC phase ($\mathcal{C}_1=\pm 1$).\par
On the other hand, it is well-known that a time-reversal invariant HgTe-QW hosts helical edge states at its boundary~\cite{Bernevig2006, Koenig2007, Qi2008} when the mass parameter $M$ is negative ($B/M>0$).

In the following, we will compare the chiral Majorana edge states and the
helical (Dirac) edge states, assuming that only the BIA-SOC term is present. \par
Chiral Majorana edge states and helical edge states at a single boundary are mutually
exclusive. Firstly, helical edge states are not topologically protected when time reversal symmetry is broken by an
Zeeman field, which is needed for the emergence of chiral
Majorana edge states~\cite{Qi2008}. Secondly, it has recently been shown that even in the
presence of time reversal symmetry, the introduction of the superconducting term $\Delta \tau_x$ adiabatically
interconnects the regions that in its absence would differ by their
topological invariants~\cite{Budich2013}. \par
In the parameter range where there is a chiral Majorana edge state, there are
therefore no additional helical edge states within the bulk energy gap,
see~\fref{fig:single_boundary_DiracMajorana} (a,b). Even if the mass gap $M$ is
negative, potential helical edge states are gapped out by the proximity of the superconductor.\par
For comparison to the chiral Majorana edge states, the helical edge states are shown in
the absence of both the Zeeman field and the superconductor in~\fref{fig:single_boundary_DiracMajorana} (c,d), where in contrast to the
standard HgTe the additional BIA term is included. As reported
in~\cite{Michetti2012b}, it has a negligible effect, and the velocity of the
helical edge states is still well-described
by~\cite{Zhou2008}
\begin{equation}
E=-\frac{D}{B}M\pm k_xA\sqrt{\frac{B^2-D^2}{B^2}}
\label{eq:helicaldispersion}
\end{equation}
for small $|k_x|$. The chiral and helical edge states differ greatly in their
velocities, which are given by~\eref{eq:MajoranaVelocity}
and~\eref{eq:helicaldispersion}, respectively. The
offset is also different: While chiral Majorana edge states are confined to the crossing
$k_x=E=0$ by particle hole symmetry, the helical edge states are only restricted to
crossing each other somewhere on the $E$ axis by time reversal symmetry.\par
Helical and chiral edge states have similar widths. While recent calculations for the width of helical edge states predict
an edge width under 100\,nm in a HgTe-QW near the Dirac point~\cite{Zhou2008},
experimentally confirmed up to a factor of 2 in~\cite{Ma2012}, the chiral
Majorana edge states typically extend to about 200\,nm-300\,nm in a HgTe-QW for the parameters
in~\tref{tab:HgTeValues}.

\begin{figure*}[htbp]
\centering
\subfigure[]{\includegraphics[width=0.32\textwidth]{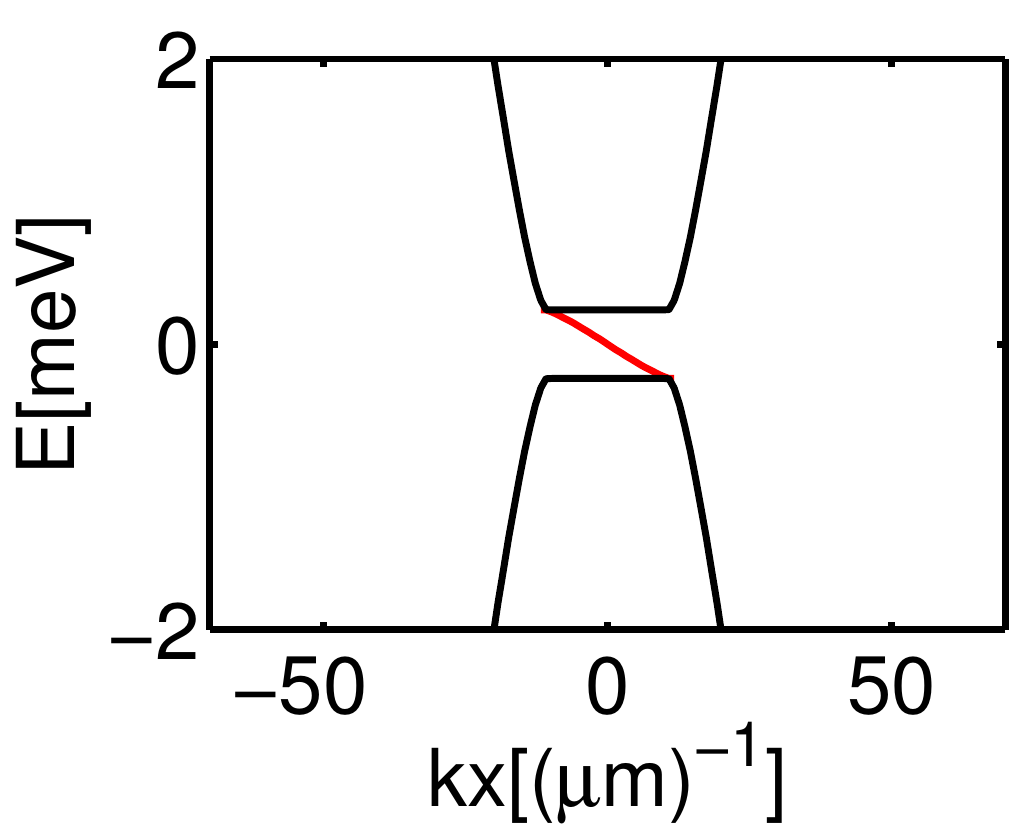}}
\subfigure[]{\includegraphics[width=0.32\textwidth]{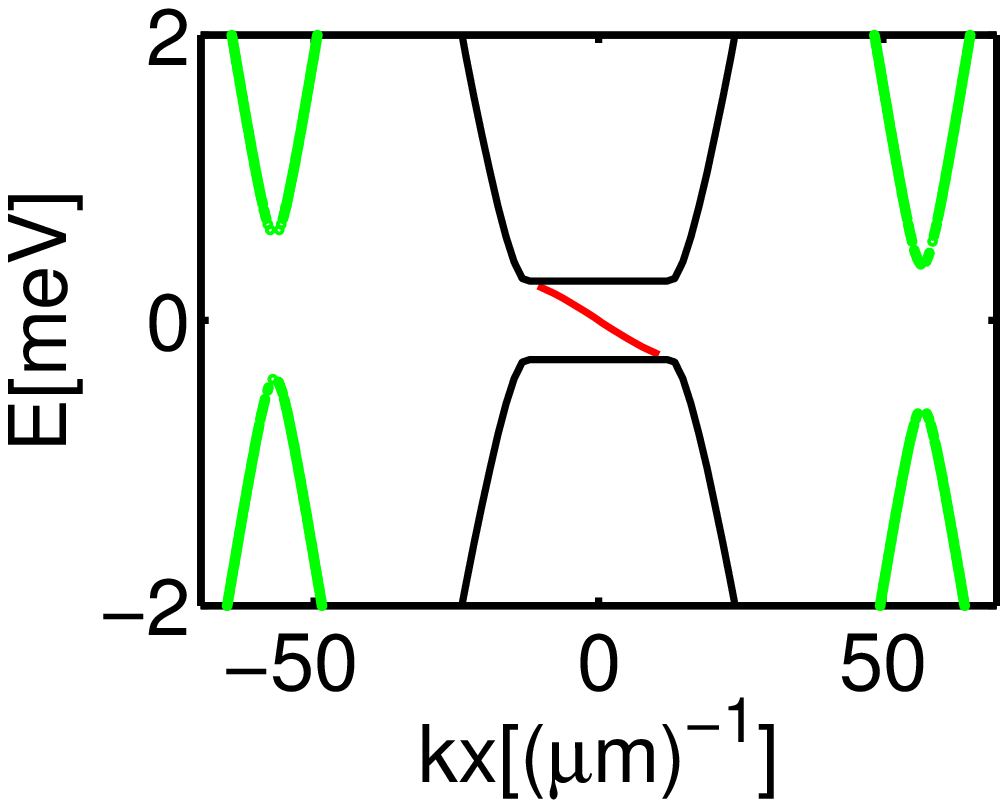}}\\
\subfigure[]{\includegraphics[width=0.32\textwidth]{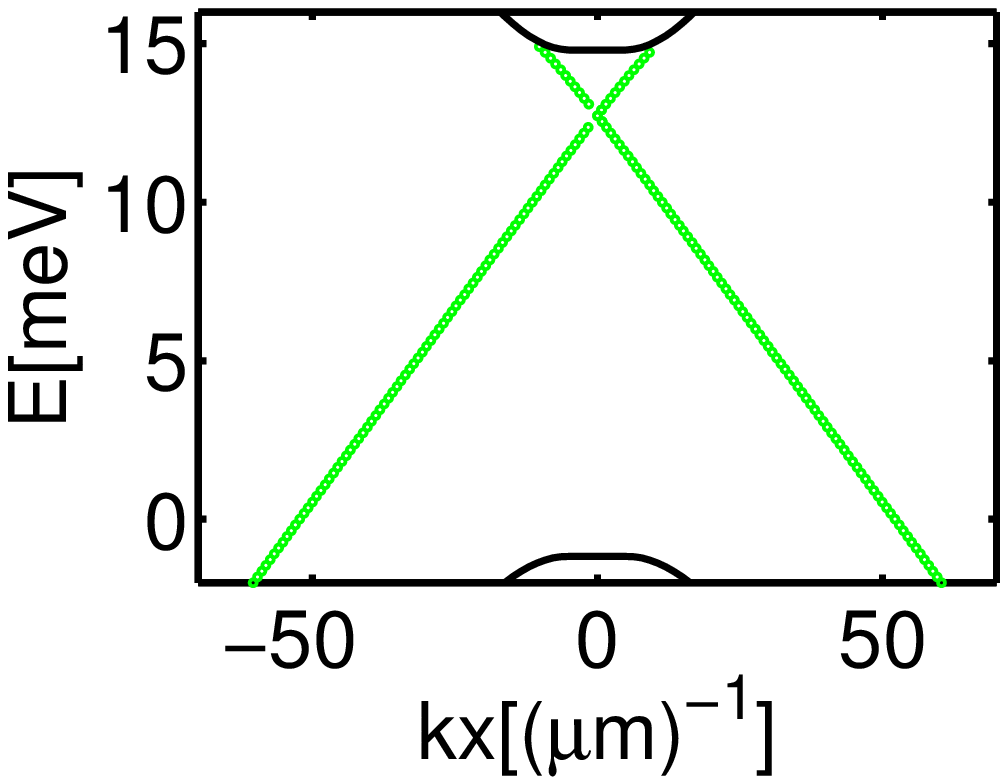}}
\subfigure[]{\includegraphics[width=0.32\textwidth]{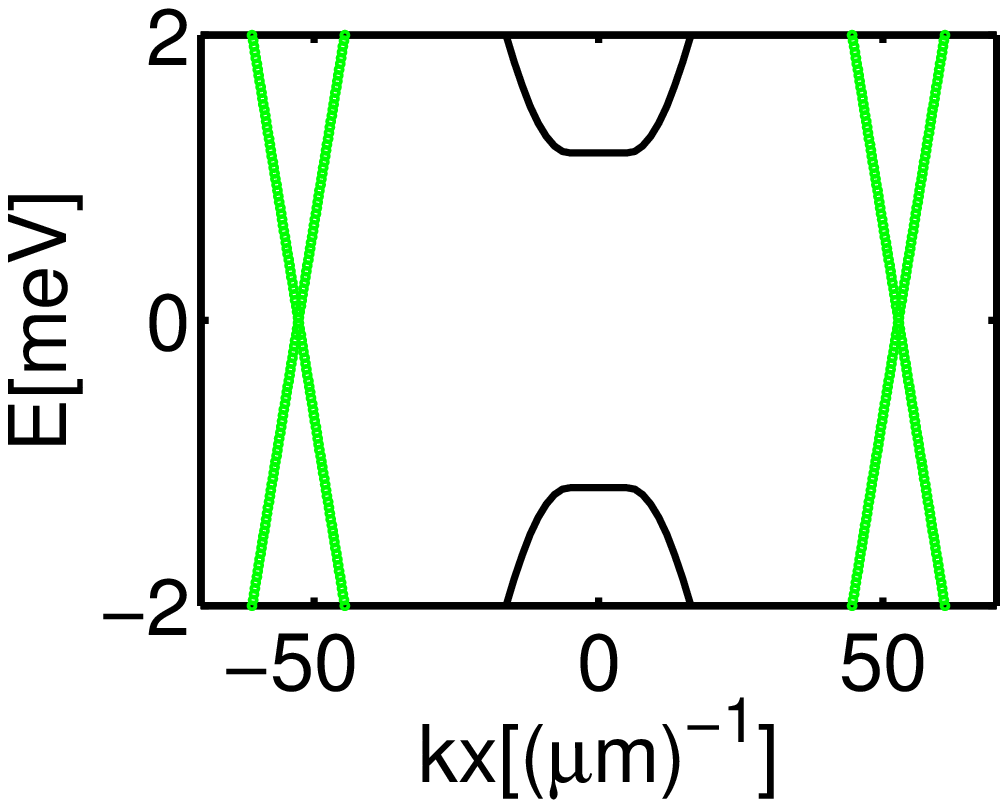}}
\caption{Energy dispersion with a hard wall boundary in $x$-direction for
  a HgTe-QW with topologically
  nontrivial parameters (a) $C=-7.8$\,meV, $M=8$\,meV and (b) $C=7.8$\,meV,
  $M=-8$\,meV. This allows chiral Majorana edge states to form. All other parameters except the Rashba-SOC, which
    is set to zero, are taken
    from~\tref{tab:HgTeValues}. In (c,d), the
  energy dispersions of a time-reversal invariant HgTe-QW in the
  inverted regime without
  the presence of the superconductor are shown ($C=6.8$\,meV, $M=-8$\,meV). For better comparison with
  (b), the degrees of freedom have been doubled in (d) by reflection at the
  Fermi level.}
\label{fig:single_boundary_DiracMajorana}
\end{figure*}

\subsection*{Summary}
In summary, we have shown that a HgTe/CdTe QW in the proximity to an s-wave superconductor and the presence of a Zeeman-splitting can represent a TSC with chiral one-dimensional edge states. At $k_x$=$E$=0, these
quasiparticles have the Majorana property of being their own
antiparticle.\par 
Importantly, an external electric field that leads to an additional Rashba-SOC term is not required, as long as the intrinsic BIA is
large enough. While in most publications, BIA is discussed as an
unwanted but small perturbation, in this case a high BIA
would be of interest. Interestingly, if Rashba-SOC and BIA-SOC are
both present, the group velocity $v=\frac{1}{\hbar}\frac{\partial E}{\partial
  k_x}$ of the chiral edge state depends crucially on the interplay between both spin-orbit
interactions and can be reversed if the Rashba-SOC is tuned from zero to a value that exceeds the bulk inversion asymmetry term. We discussed in analytical and numerical terms in which parameter regime the TSC exists using the extended four-band model of BHZ valid near the topological phase transition between a trivial and a non-trivial TI that would exist without the superconducting and Zeeman terms.\par  
 Although many systems hosting chiral Majorana fermions have so far been
presented~\cite{Fu2008, Oreg2010}, this is the first proposal to switch between one-dimensional Majorana- and Dirac edge states
in a (HgTe) QW, rather than the surface of a 3D TI, where different scenarios to convert between
one-dimensional Majorana- and Dirac edge states have previously been
proposed~\cite{Fu2009a, Akhmerov2009, Liu2011}, and the first time chiral Majorana- and helical Dirac
edge states appear in the same system.   \par
The possibility to switch between chiral Majorana- and
helical Dirac edge states on the same one-dimensional boundary might not only enhance
experimental control in the detection of Majorana quasiparticles, but could also provide an interesting
framework that combines dissipationless transport and the non-local storage
of quantum information. \par
Importantly, we would like to point out that our analysis should equally well
hold for other 2D TIs, like type-II QWs made of
InAs/GaSb/AlSb, which have intrinsically inverted
bandstructures and similar inversion symmetry breaking SOC terms~\cite{Liu2008a}. Experimental evidence of the quantum spin Hall
effect in these systems has recently been given~\cite{Knez2011}, also in
connection with superconducting contacts~\cite{Knez2012}. Further, it has been
proposed that bilayer HgTe-based QWs in the presence of an interlayer voltage
allows for an all electrically tunable TI~\cite{Michetti2012a, Michetti2013}.

\subsection*{Acknowledgments}
We would like to thank Hartmut Buhmann for enlightening  discussions, Jan Budich and Bj\"orn Trauzettel for useful comments on the manuscript and financial support from the "NTH school for contacts in nanosystems" and the DFG grant RE 2978/1-1.
 
\clearpage 
\appendix
\section*{Appendix}
\setcounter{section}{1}
If the mass gap $M$ is the relevant energy scale of the problem, leading to
only four relevant $E1$-like energy bands, this allows us to reduce the
eight-band Hamiltonian~\eref{eq:fullHamiltonian} onto an effective four-band Hamiltonian.\par  
To this end, we divide the Hamiltonian into three parts, 
\begin{eqnarray}
H=H_0+H_1+H_2\label{eq:pertHamiltonian}\\
H_0=\left(C+Ms_z\right)\tau_z\\
\eqalign{H_1=R_0\frac{\left(s_z+1\right)}{2}(\hat{k}_x\sigma_y-\hat{k}_y\sigma_x)\tau_z
\\
\quad+\left(\Delta_++\Delta_-s_z\right)\tau_x - \left(B+D\right)\hat{k}^2+\left(B_++B_-s_z\right)\sigma_z}\\
H_2=A\hat{k}_xs_x\sigma_z\tau_z-A\hat{k}_ys_y\tau_z +h\sigma_ys_y\tau_z .
\end{eqnarray}
where for clarity, in contrast to the more general
Hamiltonian~\eref{eq:fullHamiltonian}, we set $\theta=0$.\par 
In the following, $H_0$ (containing only $s_z$) is considered as the
unperturbed Hamiltonian and $H_1$ (containing only $s_z$)
and $H_2$ (containing $s_y$) as the block-diagonal and block-offdiagonal perturbations,
respectively. The perturbations are small in the sense that the mass gap M is
the dominant energy scale, i.e.
\begin{equation}
|\Delta_\pm|, |B_\pm|,|h|,|A||k|,|R_0||k|,|B||k|^2,|D||k|^2 \ll |M|
\label{eq:assumption}
\end{equation}  
In order to approximately diagonalize the Hamiltonian $H$, 
\begin{equation}
\tilde{H}=e^{-S}He^S
\end{equation}
we apply quasi-degenerate perturbation theory (``L\"owdin partitioning'')~\cite{Winkler2003}.\par
The important matrix for the transformation, $S$, consists of different orders
of the perturbation parameter, $S=S^{(1)}+\mathcal{O}(2)$. It can be
shown that  
\begin{equation}
\tilde{H}=H_0+H_1+\frac{1}{2}[H_2,S^{(1)}]+\mathcal{O}(3)
\end{equation}
is block-diagonal up to second order if $S^{(1)}$ is determined in such a way that
\begin{equation}
[H_0,S^{(1)}]=-H_2.
\end{equation}
For the Hamiltonian~\eref{eq:pertHamiltonian} it is straight-forward to show\footnote{under the requirement
  that the parameters do not depend on $x$ or $y$.} that  
\begin{equation}
S^{(1)}=+i\frac{h}{2M}s_x\sigma_y -i\frac{A\hat{k}_x}{2M}s_y\sigma_z-i\frac{A\hat{k}_y}{2M}s_x
\end{equation}
\begin{equation}
\eqalign{
\tilde{H}=\mu(s_z,\hat{k})\tau_z+B(s_z)\sigma_z+\Delta(s_z)\tau_x \\
+\tilde{h}\hat{k}_x\sigma_x\tau_z-\tilde{h}\hat{k}_ys_z\sigma_y\tau_z +R(s_z)\hat{k}_x\sigma_y\tau_z-R(s_z)\hat{k}_y\sigma_x\tau_z
}
\label{eq:hamiltoniantransformed}
\end{equation}
where the new parameters 
\begin{eqnarray}
\mu(s_z,\hat{k}^2)=\epsilon(\hat{k}_y) +\left(M(\hat{k}^2)+\frac{A^2\hat{k}^2}{2M}+\frac{h^2}{2M} \right)s_z \\
B(s_z)=B_+ +B_-s_z\\
\Delta(s_z)=\Delta_++\Delta_-s_z \\
\tilde{h}=\frac{Ah}{M}\\
R(s_z)=R_0\frac{\left(s_z+1\right)}{2}\label{eq:parameterabs}
\end{eqnarray}
have been introduced.
The advantage of the transformed Hamiltonian~\eref{eq:hamiltoniantransformed}
is that $s_z$ is (approximately) a good quantum number. For further
calculations, we neglect the $s_z=-1$ solutions in the main
text~\eref{eq:effectivehamiltonian}, and retrieve the angular dependence on
$\theta$ by the simple replacement rules
\begin{eqnarray}
h\sigma_y\rightarrow h\left(
  \cos{2\theta}\sigma_y+\sin{2\theta}\sigma_x\right)\\
h\sigma_x\rightarrow h\left(
  \cos{2\theta}\sigma_x-\sin{2\theta}\sigma_y\right)
\end{eqnarray}
which amount to a rotation in the $xy$-plane.

\end{document}